\documentclass[a4paper,fleqn,usenatbib]{mnras}

\usepackage[T1]{fontenc}
\usepackage{ae,aecompl}
\usepackage{graphicx,color}   
\usepackage{amsmath}    
\usepackage{amssymb}    
\usepackage{ascmac}
\usepackage{fancyvrb}
\usepackage{fancybox}

\newcommand{\hatx}{\hat{\bfx}}
\newcommand{\hats}{\hat{\bfs}}
\newcommand{\hatq}{\hat{\bfq}}
\newcommand{\hatz}{\hat{\bfz}}
\newcommand{\hatd}{\hat{\bfd}}
\newcommand{\hatk}{\hat{\bfk}}

\newcommand{\bfzero}{\mbox{\boldmath$0$}}
\newcommand{\bfd}{\mbox{\boldmath$d$}}
\newcommand{\bfv}{\mbox{\boldmath$v$}}

\newcommand{\bfx}{\mbox{\boldmath$x$}}
\newcommand{\bfk}{\mbox{\boldmath$k$}}
\newcommand{\bfp}{\mbox{\boldmath$p$}}
\newcommand{\bfq}{\mbox{\boldmath$q$}}

\newcommand{\bfs}{\mbox{\boldmath$s$}}

\newcommand{\bfz}{\mbox{\boldmath$z$}}

\newcommand{\bfA}{\mbox{\boldmath$A$}}
\newcommand{\bfB}{\mbox{\boldmath$B$}}

\newcommand{\bfPsi}{\mbox{\boldmath$\Psi$}}
\newcommand{\bfK}{\mbox{\boldmath$K$}}
\newcommand{\bfS}{\mbox{\boldmath$S$}}

\newcommand{\bfU}{\mbox{\boldmath$U$}}

\newcommand{\bfQ}{\mbox{\boldmath$Q$}}
\newcommand{\bLX}{b_{\rm X}^{\rm L}}
\newcommand{\bLY}{b_{\rm Y}^{\rm L}}
\newcommand{\DD}{D_{\rm\scriptscriptstyle X}D_{\rm\scriptscriptstyle Y}}
\newcommand{\RX}{R_{\rm\scriptscriptstyle X}}
\newcommand{\RY}{R_{\rm\scriptscriptstyle Y}}
\newcommand{\curlA}{\mbox{\boldmath$\mathcal{A}$}}

\newcommand{\deltalin}{\delta_{\rm L}}

\title[Wide-angle effects at quasi-linear scales]{Wide-angle redshift-space distortions at quasi-linear scales: cross-correlation functions from Zel'dovich approximation}
\author[A. Taruya et al.]{
\parbox{\textwidth}{
Atsushi Taruya$^{1,2}$, Shohei Saga$^{1}$, Michel-Andr\`es Breton$^{3}$, Yann Rasera$^{4}$, Tomohiro Fujita$^{5,6}$
}
\vspace*{15pt} \\
$^{1}$Center for Gravitational Physics, Yukawa Institute for Theoretical Physics, Kyoto University, Kyoto 606-8502, Japan
\\
$^{2}$Kavli Institute for the Physics and Mathematics of the Universe (WPI), The University of Tokyo Institutes for Advanced Study, \\
The University of Tokyo, 5-1-5 Kashiwanoha, Kashiwa, Chiba 277-8583, Japan
\\
$^{3}$Aix Marseille Univ, CNRS, CNES, LAM, Marseille, France
\\
$^{4}$LUTH, Observatoire de Paris, PSL Research University, CNRS, Universit\'e Paris Diderot, Sorbonne Paris Cit\'e 5 place Jules Janssen, 
\\
F-92195 Meudon, France
\\
$^{5}$Department of Physics, Kyoto University, Kyoto, 606-8502, Japan
\\
$^{6}$D\'epartment de Physique Th\'eorique and Center for Astroparticle Physics,
Universit\'e de Gen\'eve, Quai E. Ansermet 24, CH-1211 \\
Gen\'eve 4, Switzerland
\\
}
\date{Accepted XXX. Received YYY; in original form ZZZ}

\pubyear{2019}

\begin{document}
\VerbatimFootnotes

\thisfancyput(14.8cm,0.5cm){\large{YITP-19-76}}

\label{firstpage}
\pagerange{\pageref{firstpage}--\pageref{lastpage}}
\maketitle
\begin{abstract}
Redshift-space distortions (RSD) in galaxy redshift surveys generally break both the isotropy and homogeneity of galaxy distribution. While the former aspect is particularly highlighted as a probe of growth of structure induced by gravity, the latter aspect, often quoted as wide-angle RSD but ignored in most of the cases, will become important and critical to account for as increasing the statistical precision in next-generation surveys. However, the impact of wide-angle RSD has been mostly studied using linear perturbation theory. In this paper, employing the Zel'dovich approximation, i.e., first-order Lagrangian perturbation theory for gravitational evolution of matter fluctuations, we present a quasi-linear treatment of wide-angle RSD, and compute the cross-correlation function. The present formalism consistently reproduces linear theory results, and can be easily extended to incorporate relativistic corrections (e.g., gravitational redshift). 
\end{abstract}


\section{Introduction}
\label{sec:intro}

The large-scale structure of the Universe, as partly seen by galaxy distributions, has evolved dominantly under the influence of gravity and cosmic expansion. While the spatial inhomogeneity of matter and galaxy distribution is in nature random and stochastic, it is supposed to be statistically homogeneous and isotropic. However, the observation can break homogeneity and isotropy. In particular, the galaxy distribution observed via spectroscopic survey appears distorted along the observer's line-of-sight due to the contribution of peculiar velocities to the measured redshift of a galaxy, referred to as the redshift-space distortions (RSD).

RSD generally complicates the data analysis and cosmological interpretation of the observed galaxy clustering, but one advantage may be that RSD provides an additional information on the velocity field at large scales. Indeed, taking the distant-observer or plane-parallel limit, the statistical homogeneity is approximately restored, and the apparent anisotropies induced by RSD is characterized well by the multipole expansion with respect to the line-of-sight direction of the distant observer. On large scales, such anisotropies are described by linear theory only with few low multipoles, which tell us that the strength of anisotropies is directly related to the growth of cosmic structure induced by gravity \citep{Kaiser1987,1992ApJ385L5H}. In this respect, the measurement of clustering anisotropies caused by RSD offers an exciting opportunity to probe gravity on cosmological scales. This explains why there have been so far numerous works in both theory and observation to model, predict, and measure the anisotropies of galaxy clustering, leading to fruitful cosmological constraints \citep[e.g.,][]{Linder:2007nu,Song:2008qt,Percival:2008sh,Taruya:2010mx,Vlah:2012ni,Carlson_Reid_White2013,Beutler_etal2014,SDSS_BOSS_DR12_2017}.

With the wealth of large data set from future galaxy surveys, the statistical precision will be substantially improved, and it will help to further tighten the cosmological constraints \citep[see][for a review]{Weinberg_etal2013}. However, one must be careful in characterizing the galaxy clustering. Since the statistical homogeneity is not fully ensured in the presence of RSD, and the techniques developed so far in both measurement and theoretical predictions heavily rely on statistical homogeneity, the impact of its violation, often quoted as wide-angle effect, can introduce systematics in constraining cosmology with RSD measurement, potentially leading to a biased cosmological result.

Indeed, the impact of wide-angle effect on RSD have been long studied in both analytical and numerical approaches, and there is thus a large number of literature on this topic, including early works \citep{Fisher_etal1994,Zaroubi_Hoffman1996,Heavens_Taylor1995,Hamilton_Culhane1996,Szalay_Matsubara_Landy1998,Matsubara2000}. For cosmological data analyses including the wide-angle effect, see e.g., \citet{Tadros_etal1999,Matsubara_etal2000,Pope_etal2004,Okumura_etal2008}.

One important consequence of the wide-angle effect is that when naively applying the multipole expansion in a certain line-of-sight definition, it produces new contributions not only at even multipoles but also at odd multipoles. Indeed, such contributions have been recently detected and measured at a statistically significant level from SDSS BOSS DR12 \citep{Beutler_Castorina_Zhang2019} \citep[see ][for the analysis using DR10]{Gaztanaga_Bonvin_Hui2017}. This immediately implies that as increasing the statistical precision, the wide-angle effect can definitely give an impact on cosmological interpretation from future surveys, and theoretical prediction and measurement technique beyond the distant-observer limit have to be developed from a modern viewpoint \citep{Yoo_Seljak2015,Castorina_White2018a,Castorina_White2018b,Beutler_Castorina_Zhang2019}.

There is also another motivation why we need to care about wide-angle effect. In general, the observed galaxy distributions are further distorted due to the relativistic corrections that arise from the light propagation in an inhomogeneous universe. For instance, a measurement of redshift receives corrections not only from galaxy's peculiar motion by Doppler effect, but also from the gravity induced by galaxy and foreground large-scale structure, i.e., gravitational redshift and integrated Sachs-Wolfe effects \citep[e.g.,][]{Yoo_etal2009, Yoo2010, Bonvin_Durrer2011, Yoo_etal2012, Challinor_Lewis2011}. Those relativistic contributions are known to produce anisotropies in the observed galaxy distributions \citep[e.g.,][]{Bertacca_etal2012,Raccanelli_etal2018}, and some of the effects can generate odd multipoles in the cross-correlation function and cross power spectrum between different biased objects \citep{McDonald2009,Bonvin_Hui_Gaztanaga2014}. Recent numerical studies taking consistently the relativistic effects into account suggest that relativistic contributions become manifest at large scales \citep{Breton_etal2019}, and could be detected in future surveys \citep[see][for a recent measurement]{Alam_etal2017}. Thus, a precision measurement of odd multipoles can offer an interesting cosmological test of general relativity, alternative to the standard RSD measurement. Nevertheless, relativistic contributions are basically tiny, and one must be careful to discriminate from the wide-angle contributions, which also produce non-vanishing odd multipoles.

In these respects, a precision theoretical modeling of RSD taking account of wide-angle effect is a rather critical issue. Beyond linear theory, however, except the numerical study using $N$-body simulations \citep[e.g.,][]{Raccanelli_etal2010}, little analytical work has been done \citep[but see][]{Shaw_Lewis2008}. Recently, \citet{Castorina_White2018b} have presented the first quasi-linear treatment of the wide-angle effects based on the Zel'dovich approximation \citep{Zeldovich1970,Novikov1969JETP,Shandarin_Zeldovich1989}, particularly focusing on the auto-correlation function. In this paper, adopting the same Zel'dovich approximation, we generalize it to the calculation of the cross-correlation function of galaxies/halos. Along the lines of generalization, we clarify similarities and differences between our formalism and that of \citet{Castorina_White2018b}, who actually considered part of the wide-angle terms with the Zel'dovich approximation. Our formalism takes into account all possible wide-angle terms relevant at the Newtonian level, assuming the uniform radial selection function. With this treatment, it is shown to be consistent with linear theory of wide-angle RSD discussed in the literature. We then study the impact of wide-angle effects in the cross-correlation functions. The cross correlation between different biased objects is known to break the symmetry of the pair counting, and in the presence of wide-angle effects, this can produce an additional contribution to the anisotropies in the two-point statistics. Comparing the Zel'dovich approximation with linear theory predictions as well as $N$-body simulations, we quantitatively investigate the possible impact of its nonlinear effect, particularly focusing on the weakly nonlinear scales. In a separate paper, on the basis of the formalism in the present paper, we will further incorporate the relativistic corrections into the prediction of cross-correlation functions, and make a detailed comparison between analytical predictions and simulations with relativistic corrections.

This paper is organized as follows. In Sec.~\ref{sec:cross_correlation}, after briefly mentioning the redshift-space distortions, we present an analytical framework to compute the cross-correlation functions at quasi-linear regime, employing the Zel'dovich approximation. Several remarks on the statistical calculation are addressed together with the comments on the treatment by \citet{Castorina_White2018b}. Then, in Sec.~\ref{sec:results}, we present the results based on our quasi-linear formalism, and quantify the nonlinear impacts of the wide-angle effects on the cross-correlation functions, which are compared with linear theory predictions and $N$-body simulations. Our important findings and an implications are summarized in Sec.~\ref{sec:conclusion}. Derivation of the analytical expressions in Zel'dovich approximation as well as the linear theory formulas for cross-correlation functions are presented in detail in Appendix \ref{appendix:derivation_DD_RX} and \ref{sec:linear_theory}, respectively, together with supplemental formulas and proof in Appendix \ref{appendix:formula_Gaussian_integrals} and \ref{appendix:recovery_correlation}.

\begin{figure}
\begin{center}
\includegraphics[width=6cm]{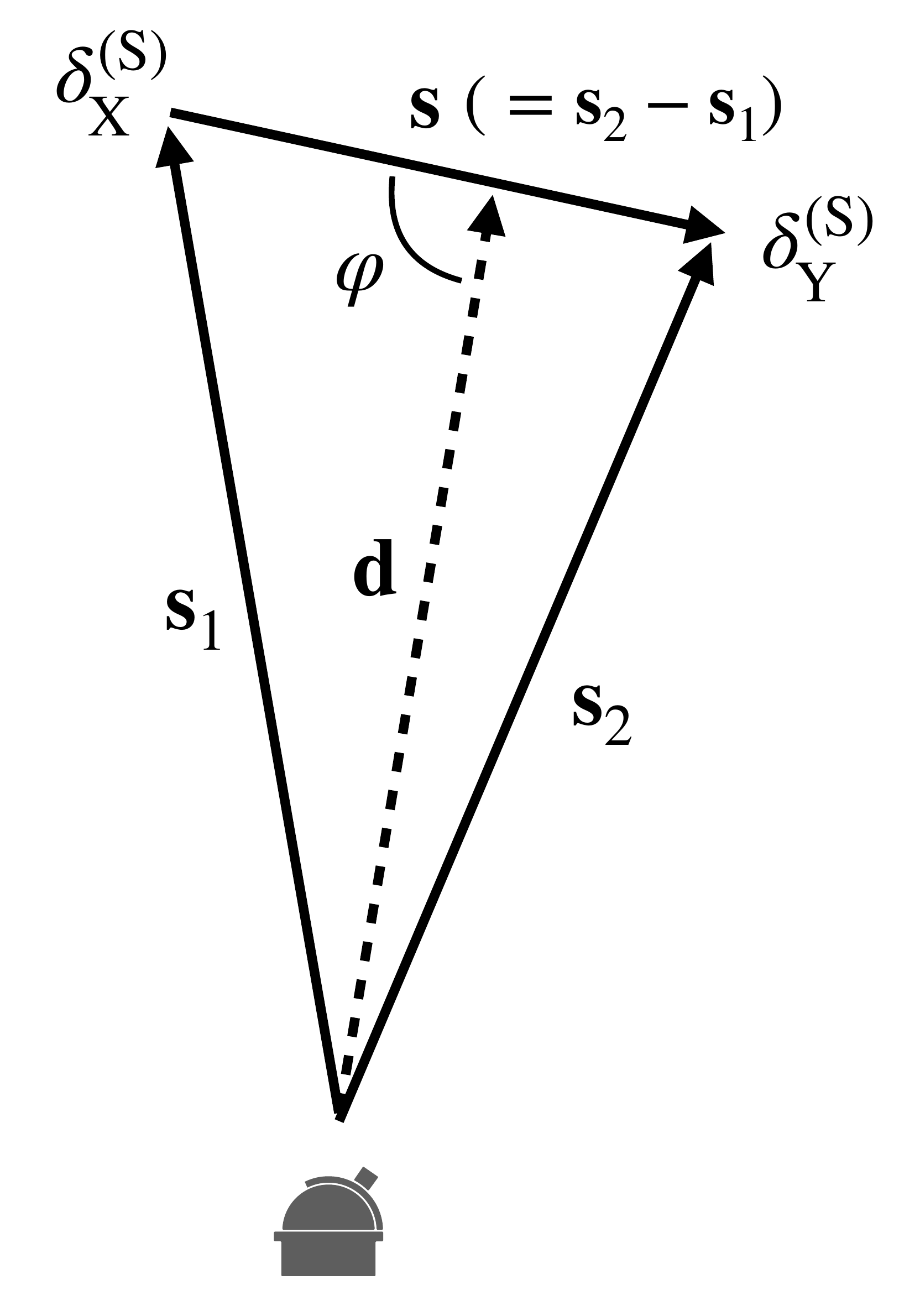} 
\end{center}
\caption{Geometric configuration of redshift-space cross-correlation function. 
Along the line-of-sight direction $\bfd$, a pair of objects $X$ and $Y$ is found at the positions $\bfs_1$ and $\bfs_2$, where the density fields,  denoted by $\delta_{\rm X}^{\rm(S)}$ and $\delta_{\rm Y}^{\rm(S)}$, is measured. The separation between these two objects is defined by $\bfs\equiv\bfs_2-\bfs_1$. Misalignment between $\bfs$ and $\bfd$ is characterized by the angle $\varphi$ or the directional cosine given by $\mu\equiv\cos\varphi$. Note that at this point, the meaning of line-of-sight direction is not well-defined, and will be later specified (see Sec.~\ref{sec:results}).  
\label{fig:geometry_xicross}}
\end{figure}

\section{Wide-angle cross-correlation function in redshift space}
\label{sec:cross_correlation}

In this paper, we are interested in computing and predicting the correlation function in redshift space without taking the distant-observer or plane-parallel limit.
Here, we only consider the Doppler effect as a major source to cause RSD. An extension to include relativistic correction will be studied in a separate paper. In the presence of Doppler effect only, the comoving position at a given redshift $z$ in redshift space, $\bfs$, is related to the real-space counterpart $\bfx$ through
\begin{align}
 \bfs=\bfx+\frac{1}{a\,H}(\bfv\cdot\hatx)\,\hatx,
\label{eq:redshift_space}
\end{align}
where $\bfv$ is the velocity field at real-space position $\bfx$, and $\hatx$ is the unit vector defined by $\hatx\equiv\bfx/|\bfx|$. The quantities $a$ and $H$ are respectively the scale factor of the Universe and Hubble parameter at a given redshift $z$. Note that in the distant-observer limit, observer's line-of-sight vector, $\hatx$, is replaced with a specific direction vector $\hatz$.

With the definition of redshift space given above,  consider the density fluctuations. Denoting the number density field of the objects $X$ by $n_{\rm X}^{\rm (S)}(\bfs)$,  we define
\begin{align}
 \delta_{\rm X}^{\rm(S)}(\bfs) =\frac{n_{\rm X}^{\rm(S)}(\bfs)}{\langle n_{\rm X}^{\rm(S)}(\bfs)\rangle}-1. 
\label{eq:def_delta_S}
\end{align}
The bracket $\langle\cdots\rangle$ stands for the ensemble average. Then, the cross-correlation function between different species $X$ and $Y$ is given by 
\begin{align}
 \xi^{\rm(S)}_{\rm XY}(\bfs_1,\bfs_2) \equiv \langle \delta_{\rm X}^{\rm(S)}(\bfs_1)\,\delta_{\rm Y}^{\rm(S)}(\bfs_2)\rangle. 
\label{eq:def_xired_XY}
\end{align}

Note that the cross-correlation function defined above is, as opposed to the one in real space, not simply described by the function of the separation between two objects. In the presence of observer's line-of-sight vector $\hatx$ in Eq.~(\ref{eq:redshift_space}), both the statistical homogeneity and isotropy of the galaxy distributions no longer hold, and we generally need three variables to characterize the correlation function in redshift space. That is, $\xi^{\rm(S)}_{\rm XY}$ is given as a function of the distances to the objects $|\bfs_1|$ and $|\bfs_2|$, and separation $s\equiv|\bfs_2-\bfs_1|$ (see Fig.~\ref{fig:geometry_xicross}). In other words, the correlation function is described with the triangle characterized by the vectors, $\bfs_1$, $\bfs_2$, and $\bfs\equiv\,\bfs_2-\bfs_1$, and it is invariant under the transformation such that the shape of this triangle remains unchanged.

\subsection{Zel'dovich approximation}
\label{subsec:Zeldovich}

Our primary interest is to develop the quasi-linear theory of wide-angle redshift-space correlation function. For this purpose, we follow \cite{Castorina_White2018b} and use the Zel'dovich approximation, which allows us to predict the position and motion of mass element, given an initial condition of density field \citep{Zeldovich1970,Novikov1969JETP,Shandarin_Zeldovich1989}. An important building block in the Zel'dovich approximation is the displacement field of each mass element, which is given as a function of Lagrangian coordinate (initial position of each mass element), $\bfq$. In what follows, we assume that the objects of our interest to measure the correlation function simply follow the velocity flow of mass distributions (i.e., no velocity bias). Denoting the displacement field by $\bfPsi(\bfq)$, the Eulerian position $\bfx$ and velocity of mass element at $\bfx$ are then expressed as
\begin{align}
 \bfx = \bfq + \bfPsi(\bfq),\qquad
 \bfv(\bfx)=a\,\frac{d\bfPsi(\bfq)}{dt}. 
\label{eq:Euler_Lagrange_mapping} 
\end{align}
The Zel'dovich approximation gives a simple analytical expression for the displacement field in terms of the (Lagrangian) linear density field $\deltalin$ as:
\begin{align}
 \nabla \cdot\bfPsi_{\rm ZA}(\bfq) =- \deltalin(\bfq). 
\label{eq:displacement_ZA} 
\end{align}
Recalling that the linear density field is related to initial density field $\delta_0$ through $\deltalin=D_+(t)\,\delta_0$ with $D_+$ being linear growth factor, 
we have
\begin{align}
 \bfv = a H\,f(t)\,\bfPsi_{\rm ZA}(\bfq).
\label{eq:velocity_ZA} 
\end{align}
Here, the function $f$ is linear growth rate defined by 
\begin{align}
 f(t)\equiv \frac{d\ln D_+(t)}{d\ln a(t)}.
\end{align}
Substituting these relations into Eq.~(\ref{eq:redshift_space}), we obtain (hereafter we omit the subscript $_{\rm ZA}$, and simply write $\bfPsi$), 
\begin{align}
 s_i &= q_i + \Bigl\{\delta_{ij}+f\,\hat{x}_i\hat{x}_j\Bigr\}\,\Psi_j(\bfq)
\nonumber
\\
&\simeq q_i + \Bigl\{\delta_{ij}+f\,\hat{q}_i\hat{q}_j\Bigr\}\,\Psi_j(\bfq).
\label{eq:mapping_s_q-space}
\end{align}
Note that the second line is valid at first-order Lagrangian perturbation theory (i.e., Zel'dovich approximation). Here, we used the Einstein summation convention. The subscripts $i$ and $j$ take values $1$, $2$ or $3$. Eq.~(\ref{eq:mapping_s_q-space}) gives a mapping relation between redshift space and Lagrangian space, and is a basis to compute statistical quantities in redshift space given the statistical properties in Lagrangian space.

\subsection{Analytical expression}
\label{subsec:analytical_expression}

Once established the relation between Eulerian- and Lagrangian-space positions, we now express the observed number density field of the population $X$, defined in redshift space, $n_{\rm X}^{\rm(S)}$, in terms of the Lagrangian-space quantities. In what follows, we assume the linear bias relation for all objects to cross correlate. Note that the extension to incorporate the nonlinear Lagrangian bias into statistical calculation has been made in the case of distant-observer or plane-parallel limit by \citet{Carlson_Reid_White2013, Wang_Reid_White2014, White2014} \citep[see also][for slightly different formalism]{Matsubara2008b,Matsubara2014}. 

Using the number conservation in each space, we have
\begin{align}
n_{\rm X}^{\rm(S)}(\bfs)\,d^3\bfs = n_{\rm X}(\bfx) d^3\bfx=\overline{n}_{\rm X}\Bigl\{1+\bLX\,\deltalin(\bfq)\,\Bigr\} \,d^3\bfq,
\label{eq:number_density_X}
\end{align}
where $\overline{n}_{\rm X}$ is the mean number density at a given redshift, and we assume it to be constant over the survey region. The quantity $\bLX$ is the Lagrangian linear bias parameter for the population $X$, which is related to the Eulerian linear bias $b_{\rm X}$ through $b_{\rm X}=1+\bLX$. Note that $\overline{n}_{\rm X}$ does not in general coincide with the mean density in redshift space, $\langle n_{\rm X}^{\rm(S)}(\bfs)\rangle$, unless we take distant-observer or plane-parallel limit. Eq.~(\ref{eq:number_density_X}) is then rewritten with
\begin{align}
 n_{\rm X}^{\rm(S)}(\bfs) & = \overline{n}_{\rm X}\,\Bigl|\frac{\partial\bfs}{\partial \bfq}\Bigr|^{-1}\,\{1+\bLX\,\deltalin(\bfq)\}
\nonumber
\\
&= \overline{n}_{\rm X}\int d^3\bfq\,\,
\delta_{\rm D}\Bigl[\bfs-\bfq-\bfPsi^{\rm(S)}(\bfq)\Bigr]\,
\{1+\bLX\,\deltalin(\bfq)\}
\nonumber
\\
&= \overline{n}_{\rm X}\int\frac{d^3\bfk}{(2\pi)^3}\,\int d^3\bfq\,
\,e^{i\,\bfk\cdot\{\bfs-\bfq-\bfPsi^{\rm(S)}(\bfq)\}}
\{1+\bLX\,\deltalin(\bfq)\},
\label{eq:n_X_Lagrangian}
\end{align}
where the quantity $\delta_{\rm D}$ is the Dirac delta function, which is re-expressed in the third line, introducing the auxiliary variable (wave vector), $\bfk$. Here, we define the redshift-space displacement field, $\bfPsi^{\rm(S)}$ [see Eq.~(\ref{eq:mapping_s_q-space})]:
\begin{align}
 \Psi_i^{\rm(S)}(\bfq) &= 
(\delta_{ij}+f\,\hat{q}_i\hat{q}_j)\,\Psi_j(\bfq)
\nonumber
\\
&\equiv R_{ij}(\hatq)\,\Psi_j(\bfq).
\label{eq:def_Rij}
\end{align}

Substituting Eq.~(\ref{eq:n_X_Lagrangian}) into the redshift-space density fluctuation given at Eq.~(\ref{eq:def_delta_S}), the cross-correlation function $\xi_{\rm XY}^{\rm(S)}$ at Eq.~(\ref{eq:def_xired_XY}) is expressed as 
\begin{align}
1+\xi_{\rm XY}^{\rm(S)}(\bfs_1,\bfs_2) &=\Bigl\langle
\bigl\{1+\delta_{\rm X}^{\rm(S)}(\bfs_1)\bigr\}\bigl\{1+\delta_{\rm Y}^{\rm(S)}(\bfs_2)\bigr\}\Bigr\rangle
\nonumber
\\
&= \frac{\DD(\bfs_1,\bfs_2)}{\RX(\bfs_1)\RY(\bfs_2)}
\label{eq:xired_cross_DD_RR}
\end{align}
with the functions given at the denominator and numerator, $R_{\rm X,Y}$ and $\DD$, respectively defined by 
\begin{align}
 R_{\rm X,Y}(\bfs) &\equiv \int\frac{d^3\bfk}{(2\pi)^3}\int d^3\bfq\,
e^{i\bfk\cdot\{\bfs-\bfq\}}
\nonumber
\\
&\times \Bigl\langle e^{-i\bfk\cdot\bfPsi^{\rm(S)}(\bfq)} \{1+b_{\rm X,Y}^{\rm L}\deltalin(\bfq)\bigr\}
\Bigr\rangle,
\label{eq:R_term}
\end{align}
\begin{align}
 \DD(\bfs_1,\bfs_2) &\equiv \int\frac{d^3\bfk_1 d^3\bfk_2}{(2\pi)^6}\int d^3\bfq_1 d^3\bfq_2\,
\nonumber
\\
&\times e^{i\bfk_1\cdot\{\bfs_1-\bfq_1\}+i\bfk_2\cdot\{\bfs_2-\bfq_2\}}
\nonumber
\\
&\times \Bigl\langle e^{-i\bfk_1\cdot\bfPsi^{\rm(S)}(\bfq_1)-i\bfk_2\cdot\bfPsi^{\rm(S)}(\bfq_2)}
\nonumber
\\
&\times \{1+\bLX\deltalin(\bfq_1)\bigr\}\bigl\{1+\bLY\deltalin(\bfq_2)\bigr\}
\Bigr\rangle.
\label{eq:DD_term}
\end{align}

Note that the ensemble average in these expressions is evaluated with respect to the randomness of linear density field $\deltalin$, which is, in Lagrangian space, statistically homogeneous and isotropic. Thus, one may expect that taking the average, quantities with brackets are expressed, after all, as function of separation only, i.e., $|\bfq_2-\bfq_1|$. If this is the case, the expressions given above can be drastically simplified under the Gaussian initial condition. Performing analytically the integrals over wavenumbers, $R_{\rm X,Y}$ is found to be $1$, and $\DD$ is finally reduced to the form involving three-dimensional Gaussian integral, which can be evaluated numerically with a better convergence \citep[e.g.,]{BondCouchman1986,Schneider_Bartelmann1995,Fisher_Nusser1996,Taylor_Hamilton1996}. However, this simplification can be applied only in the distant-observer or plane-parallel limit. Due to the position-dependent matrix $R_{ij}$ in the displacement field $\bfPsi^{\rm(S)}$, the brackets have non-trivial dependence of the Lagrangian positions even after taking the averages. This is solely due to the wide-angle RSD that observer's line-of-sight direction varies over the sky, and cannot be taken to be a specific direction.

Thus, taking a proper account of the wide-angle effect, the calculation of Eqs.~(\ref{eq:R_term}) and (\ref{eq:DD_term}) ceases to be trivial. Nevertheless, it is still possible to reduce the expressions of $R_{\rm X,Y}$ and $\DD$ given above to those involving three- and six-dimensional Gaussian integrals, respectively. In Appendix \ref{appendix:derivation_DD_RX}, we derive the final forms. The expression of $R_{\rm X,Y}$ is summarized as follows: 
\begin{align}
 \RX(\bfs) = \RY(\bfs) = \int\frac{d^3\bfq}{(2\pi)^{3/2}|\mbox{det}\mbox{\boldmath$A$}|} \,
e^{-(1/2)A^{-1}_{ij}(s-q)_i(s-q)_j},
\label{eq:expression_R_part}
\end{align}
where the matrix $A_{ij}$ is defined by
\begin{align}
 A_{ij}(\bfq) &\equiv \langle \Psi_i^{\rm(S)}(\bfq)\Psi_j^{\rm(S)}(\bfq)\rangle.
\label{eq:def_matrix_Aij}
\end{align}
Note that starting with the expression given at Eq.~(\ref{eq:R_X_6D-integral_form}), one can also derive an approximate expression in the following analytical form (see Appendix \ref{subsec:mean_dens}): 
\begin{align}
 \RX(s)&\simeq 1 +(2f+f^2)\Bigl(\frac{\sigma_{\rm d}^2}{s^2}+\frac{\sigma_{\rm d}^4}{s^4}+3\frac{\sigma_{\rm d}^6}{s^6}+\cdots\Bigr),  
\label{eq:R_X_approx}
\end{align}
which is accurate for the large-distance case with $\sigma_{\rm d}/s\ll1$. 
Here, $\sigma_{\rm d}$ is the rms of the Lagrangian displacement field, and its explicit expression is given at Eq.~(\ref{eq:sigma_d}).

On the other hand, for $\DD$, we introduce the six-dimensional vectors for Lagrangian and redshift-space positions, $\bfQ$ and $\bfS$, and write these as $\bfQ=(\bfq_1,\bfq_2)$ and  $\bfS=(\bfs_1,\bfs_2)$. Then, the correlation term $\DD$ is expressed as follows:
\begin{align}
&\DD(\bfs_1,\bfs_2) = \int \frac{d^6\bfQ}{(2\pi)^3|\mbox{det}\,\curlA|^{1/2}}\,e^{-(1/2)\mathcal{A}_{ab}^{-1}(S-Q)_a(S-Q)_b}
\nonumber
\\
&\qquad\quad\times \Bigl[1+\bLX \bLY\,\xi_{\rm L}(|\bfq_2-\bfq_1|)
-\mathcal{A}^{-1}_{cd}\,\mathcal{U}_c(S-Q)_d 
\nonumber
\\
&\qquad\quad-\Bigl\{ \mathcal{A}^{-1}_{cd} - \mathcal{A}^{-1}_{ce}\mathcal{A}^{-1}_{df} 
(S-Q)_e (S-Q)_f \Bigr\} \mathcal{W}_{cd}
\Bigr].
\label{eq:expression_DD_part}
\end{align}
The subscripts $a,b,\cdots$ run over $1-6$. The explicit expressions for the quantities given above, $\mathcal{U}_a$, $\mathcal{A}_{ab}$, and $\mathcal{W}_{ab}$, as well as the $3\times3$ matrix $A_{ij}$, are all presented in Appendix \ref{subsec:cross_corr}. Note that the function $\DD$ depends on the bias parameters not only explicitly in the coefficient of $\xi_{\rm L}$ but also implicitly through the definitions of $\mathcal{U}_a$ and $\mathcal{W}_{ab}$ [see Eqs.~(\ref{eq:6D_vector_U}) and (\ref{eq:6x6_matrix_W})].

In what follows, we use Eqs.~(\ref{eq:expression_R_part}) and (\ref{eq:expression_DD_part}) to give quantitative predictions of wide-angle cross-correlation function $\xi_{\rm XY}^{\rm(S)}$. Numerical integrals involved in these expressions are performed specifically with {\tt cuhre} routine in the {\tt CUBA} library \citep{Hahn:2004fe}\footnote{http://www.feynarts.de/cuba/}. 


\subsection{Relation to Castorina \& White (2018b)}
\label{subsec:Castorina_and_White}

Before closing this section, we compare our formalism in Sec.~\ref{sec:cross_correlation} with the one given in \cite{Castorina_White2018b}, who have first presented the analytical calculation of wide-angle effects beyond linear theory prediction based on the same Zel'dovich approximation as we adopted. To be precise, \cite{Castorina_White2018b} considered part of the wide-angle effects. Here, we clarify the differences between our and their treatments. 

Let us first check that the present formalism correctly reproduces the well-known linear theory result with wide-angle corrections under the uniform radial selection function. We derive the linear-order expression for the redshift-space density fluctuation, $\delta^{\rm(S)}_{\rm X}$, given at Eq.~(\ref{eq:def_delta_S}). Substituting the expression for the number density field at Eq.~(\ref{eq:n_X_Lagrangian}) into Eq.~(\ref{eq:def_delta_S}), we Taylor-expand the exponents. At the leading order, the denominator $\langle n^{\rm(S)}_{\rm X}\rangle$ does not play any role, and the expansion of the numerator leads to 
\begin{align}
 \delta_{\rm X,lin}^{\rm(S)}(\bfs)&=\int\frac{d^3\bfk}{(2\pi)^3}\,\int d^3\bfq\,
e^{i\,\bfk\cdot(\bfs-\bfq)}\,
\nonumber
\\
&\quad\qquad\qquad\qquad\times
\Bigl[\bLX\,\deltalin(\bfq)-i\,\bfk\cdot\bfPsi^{\rm(S)}(\bfq)\Bigr].
\label{eq:deltas_lin1}
\end{align}
Recalling that $\bfPsi^{\rm(S)}$ is the displacement field defined in redshift space, and it is related to the real-space displacement field through $\Psi_i^{\rm(S)}=(\delta_{ij}+f\hat{q}_i\hat{q}_j)\,\Psi_j$ [see Eq.~(\ref{eq:def_Rij})], we have
\begin{align}
 \delta_{\rm X,lin}^{\rm(S)}(\bfs)& =
\int\frac{d^3\bfk}{(2\pi)^3}\,\int d^3\bfq\,e^{i\,\bfk\cdot(\bfs-\bfq)}
\nonumber
\\
&\quad\times\,
\Bigl[\bLX\,\deltalin(\bfq)-i\,\bfk\cdot\bfPsi(\bfq)-i\,f(\bfk\cdot\hatq)\{\hatq\cdot\bfPsi(\bfq)\}\,\Bigr],
\label{eq:deltas_lin1'}
\end{align}
which can be recast as
\begin{align}
\delta_{\rm X,lin}^{\rm(S)}(\bfs)
&=\bLX\,\deltalin(\bfs)-\nabla_s\cdot\bfPsi(\bfs)-f\,\nabla_s\cdot\Bigl[\{\hats\cdot\bfPsi(\bfs)\}\,\hats\Bigr]. 
\label{eq:deltas_lin2}
\end{align}
Here, $\hats$ is unit vector given by $\hats\equiv\bfs/|\bfs|$, and the operator $\nabla_s$ stands for the divergence in the redshift-space coordinates. Note that the second and third terms at right-hand-side have been derived from 
Eq.~(\ref{eq:deltas_lin1'}) by rewriting the factor $i\,\bfk\,e^{i\,\bfk\cdot(\bfs-\bfq)}$ in the integrand  with $-\nabla_q\,e^{i\,\bfk\cdot(\bfs-\bfq)}$, and performing the integral over $\bfk$. Using the formulas $\partial \hat{s}_i/\partial s_j=(\delta_{ij}-\hat{s}_i\hat{s}_j)/|\bfs|$ and $(\hats\cdot\nabla_s)\,\hats=0$, the last term of the above expression is rewritten with 
\begin{align}
\nabla_s\cdot\Bigl[\{\hats\cdot\bfPsi(\bfs)\}\,\hats\Bigr] 
=\Bigl\{\frac{2}{s} +(\hats\cdot\nabla_s)\Bigr\}\Bigl\{\bfPsi(\bfs)\cdot\hats\Bigr\}. 
\label{eq:nabla_term}
\end{align}
Further, in Zel'dovich approximation, the real-space displacement field and its spatial derivative are related to the velocity and density field through $\bfv=a\,H\,f\,\bfPsi$ and $\nabla_s\cdot\Psi=-\deltalin$ [see Eqs.~(\ref{eq:displacement_ZA}) and (\ref{eq:velocity_ZA})]. Then, Eq.~(\ref{eq:deltas_lin2}) is finally reduced to the following form: 
\begin{align}
\delta_{\rm X,lin}^{\rm(S)}(\bfs)=b_{\rm X}\,\deltalin(\bfs)-\frac{1}{a\,H}\Bigl\{\frac{2}{s}+(\hats\cdot\nabla_s)\Bigr\}(\bfv\cdot\hats), 
\label{eq:deltas_wide-angle}
\end{align}
where the factor $1+\bLX$ has been replaced with the Eulerian linear bias parameter $b_{\rm X}$. Eq.~(\ref{eq:deltas_wide-angle}) coincides with the well-known result for redshift-space linear density field taking account of the wide-angle effects, assuming a uniform radial selection function \citep[e.g.,][]{Kaiser1987,Szalay_Matsubara_Landy1998,Yoo_Seljak2015}\footnote{To be precise, we assume the constant mean number density, and the contribution from its evolution is ignored in Eq.~(\ref{eq:deltas_wide-angle}).}. Note that the term proportional to $(2/s)\,(\bfv\cdot\hats)$ is often called the selection function terms, and in general cases with non-uniform radial selection function,  the factor $2$ is replaced with $\alpha(r)=2+d\ln\phi(r)/d\ln r$, with $\phi(r)$ being the radial selection function slowly varying function of the radial distance, $r$ \citep[e.g.,][]{Kaiser1987,Szalay_Matsubara_Landy1998,Yoo_Seljak2015,Castorina_White2018a}.

In Appendix \ref{appendix:recovery_correlation}, for the sake of the completeness, we also show that our formalism, starting from the expression given at Eq.~(\ref{eq:xired_cross_DD_RR}), consistently reproduces the linear cross-correlation function with wide-angle effects. Note that in this case, not only the numerator in Eq.~(\ref{eq:xired_cross_DD_RR}) but also the denominator, i.e., product of mean density, $\RX\RY$, play a role, and have to be taken into consideration properly.

Let us next look at the linear density field based on the treatment by \cite{Castorina_White2018b}. A crucial assumption or proposition is to rewrite the redshift-space displacement field $\bfPsi^{\rm(S)}$, given at Eq.~(\ref{eq:def_Rij}), with
\begin{align}
 ^{\rm CW}\Psi_i^{\rm(S)}(\bfq) = (\delta_{ij} + f\,\hat{s}_i\hat{s}_j)\,\Psi_j(\bfq),
\label{eq:displacement_CW}
\end{align}
where the redshift-position $\bfs$ is linked to the Lagrangian counterpart $\bfq$ through Eq.~(\ref{eq:mapping_s_q-space}). Seemingly, Eq.~(\ref{eq:displacement_CW}) is relevant, and looks equivalent to Eq.~(\ref{eq:def_Rij}) at linear order. However, substituting it into (\ref{eq:deltas_lin1}) and repeating the same calculation as given above, we obtain the following expression\footnote{When integrating Eq.~(\ref{eq:deltas_lin1}) over $\bfq$, we treat the unit vector $\hats$ in Eq.~(\ref{eq:displacement_CW}) independent of the Lagrangian position $\bfq$, in a similar way to what we derived Eq.~(\ref{eq:deltas_lin2})}: 
\begin{align}
^{\rm CW}\delta_{\rm X,lin}^{\rm(S)}(\bfs)=\bLX\,\deltalin(\bfq)-\nabla_s\cdot\bfPsi(\bfs)-f\,\Bigl[\hats\cdot \{(\hats\cdot\nabla_s)\bfPsi(\bfs)\} \Bigr], 
\label{eq:deltas_CW18_before}
\end{align}
which is finally reduced to 
\begin{align}
^{\rm CW}\delta_{\rm X,lin}^{\rm(S)}(\bfs)= b_{\rm X}\,\deltalin(\bfq)-\frac{1}{a\,H}\,(\hats\cdot\nabla_s)(\bfv\cdot\hats). 
\label{eq:deltas_CW18}
\end{align}
Compared to Eq.~(\ref{eq:deltas_wide-angle}), the above expression misses the selection function terms, proportional to $(2/s)\,(\bfv\cdot\hats)$. The second term at right-hand-side of Eq.~(\ref{eq:deltas_CW18}) is known to produce the wide-angle effect, and is shown to play an important role \citep[e.g.,][]{Bonvin_Hui_Gaztanaga2014}. On the other hand, due to the suppression factor, the contribution of the selection function terms seemingly becomes unimportant for the clustering signal at high-$z$. We will come back to this point, and discuss its actual impact in comparison with $N$-body simulations in Sec.~\ref{subsubsec:impact_selection_func}.

Another notable difference between the present paper and \cite{Castorina_White2018b} appears in the expression of redshift-space correlation function. In the present paper, starting from the mapping formula given at Eq.~(\ref{eq:redshift_space}) and (\ref{eq:mapping_s_q-space}), the redshift-space correlation function has been derived from scratch, the resultant expression of which involves six-dimensional integrals. On the other hand, \cite{Castorina_White2018b} have obtained the expression based on the real-space correlation function, replacing simply the displacement field in real space with redshift-space counterpart, $\bfPsi^{\rm(S)}$, adopting Eq.~(\ref{eq:displacement_CW}). Since their derivation makes use of the statistical isotropy and homogeneity that hold in real-space correlation function, the final expression involves only the three-dimensional integral. While this treatment greatly reduces the computational cost, translational invariance is, taking account of the wide-angle effect, violated in the actual redshift space, and thus their final expression is only valid if we take the plane-parallel limit. Despite of this, the predicted behaviors of the wide-angle corrections are found to be similar to those obtained from our treatment, especially for even multipole moments of correlation functions. We will see quantitatively the similarities and differences between their treatment and the present formalism.

\section{Results}
\label{sec:results}

In this section, based on the expression given at Eq.~(\ref{eq:xired_cross_DD_RR}) with (\ref{eq:expression_R_part}) and (\ref{eq:expression_DD_part}), we present the predictions of redshift-space cross correlation functions, which are compared with linear theory and $N$-body simulations. In doing so, we made a slight extension of the linear theory formalism to predict the cross-correlation function including wide-angle corrections. In Appendix \ref{sec:linear_theory}, we present the analytical expressions for linear cross-correlation function, and the results are summarized as the trigonometric polynomial expansion, using the technique developed by \cite{Szapudi2004} and \cite{Papai_Szapudi2008}\footnote{To be strict, the analytical expressions for linear cross-correlation function has been presented in \citet{Bonvin_Hui_Gaztanaga2014}, in a mixture of relativistic and standard RSD contributions. In Appendix \ref{sec:linear_theory}, we re-derived the full analytical expressions, leaving only the relevant standard RSD contributions, and present the results together with the formulas for multipole expansion based on three different representation of line-of-sight direction.}.

As we discussed, the cross-correlation function $\xi^{\rm(S)}_{\rm XY}$ is given as function of the three variables, associated with the triangle formed with the positions of pair of objects ($\bfs_1$ and $\bfs_2$) and their separation $\bfs\equiv \bfs_2-\bfs_1$. To characterize it, as shown in Fig.~\ref{fig:geometry_xicross}, we may introduce the line-of-sight distance $\bfd$, the vector pointing to a pair of objects from the observer, and the misalignment angle $\varphi$ between the line-of-sight direction and separation for a pair of objects. Then, the correlation function can be expressed as the function of $s=|\bfs|$, $d=|\bfd|$, and $\mu\equiv\cos\varphi=\hatd\cdot\hats$, i.e., $\xi^{\rm(S)}_{\rm XY}(s,d,\mu)$. It is convenient to express it as multipole expansion: 
\begin{align}
 \xi_{\rm XY}^{\rm(S)}(\bfs_1,\bfs_2)=\sum_\ell \xi_\ell^{\rm(S)}(s,\,d)\,\,\mathcal{P}_\ell(\mu).
\label{eq:xired_multipole}
\end{align}
Note that the multipole moment $\xi_\ell^{\rm(S)}$ depends not only on separation $s$ but also on the line-of-sight distance $d$. One thus has to further expand $\xi_\ell^{\rm(S)}$ in powers of $(s/d)$: 
\begin{align}
 \xi_\ell^{\rm(S)}(s,\,d)=\sum_n \left(\frac{s}{d}\right)^n\,\xi_{\ell,n}^{\rm(S)}(s).
\label{eq:xired_multipole_wideangle}
\end{align}
The leading-order contributions with $n=0$, i.e., $\xi_{\ell,0}^{\rm(S)}$ represent the conventional multipole correlation functions in the plane-parallel limit, where even multipoles are only relevant non-vanishing quantities. On the other hand, higher-order terms of $n\geq1$ basically describe the wide-angle corrections, for which both even and odd multipoles become generally non-zero.

\begin{figure*}
\includegraphics[width=18cm]{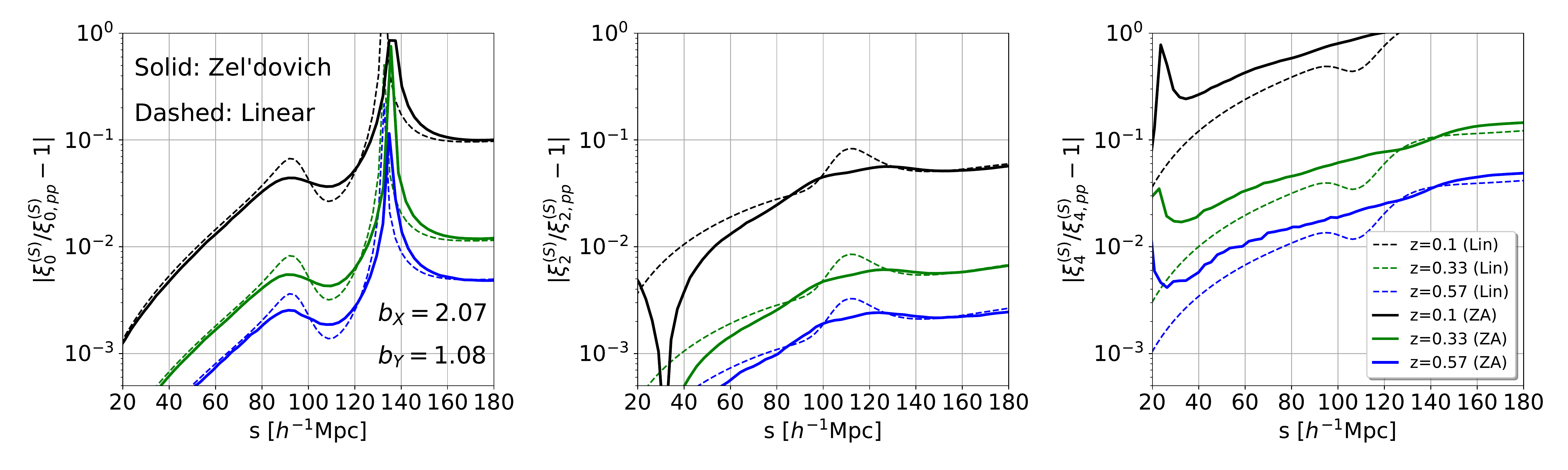}
\caption{Fractional difference of the monopole (left), quadrupole (middle), and hexadecapole (right) moments of correlation function between predictions with and without wide-angle effects, $|\xi_\ell^{\rm(S)}/\xi_{\ell,{\rm\scriptscriptstyle pp}}^{\rm(S)}-1|$, where $\xi_{\ell,{\rm\scriptscriptstyle pp}}^{\rm(S)}$ represents the multipole correlation function in the plane-parallel limit. The results at $z=0.1$, $0.33$, and $0.57$ are shown in different colors. Solid and dashed lines are respectively the predictions based on Zel'dovich approximation and linear theory, assuming the Eulerian linear bias of $b_{\rm X}=2.07$ and $b_{\rm Y}=1.08$.
\label{fig:xired024_zred}}
\includegraphics[width=12cm]{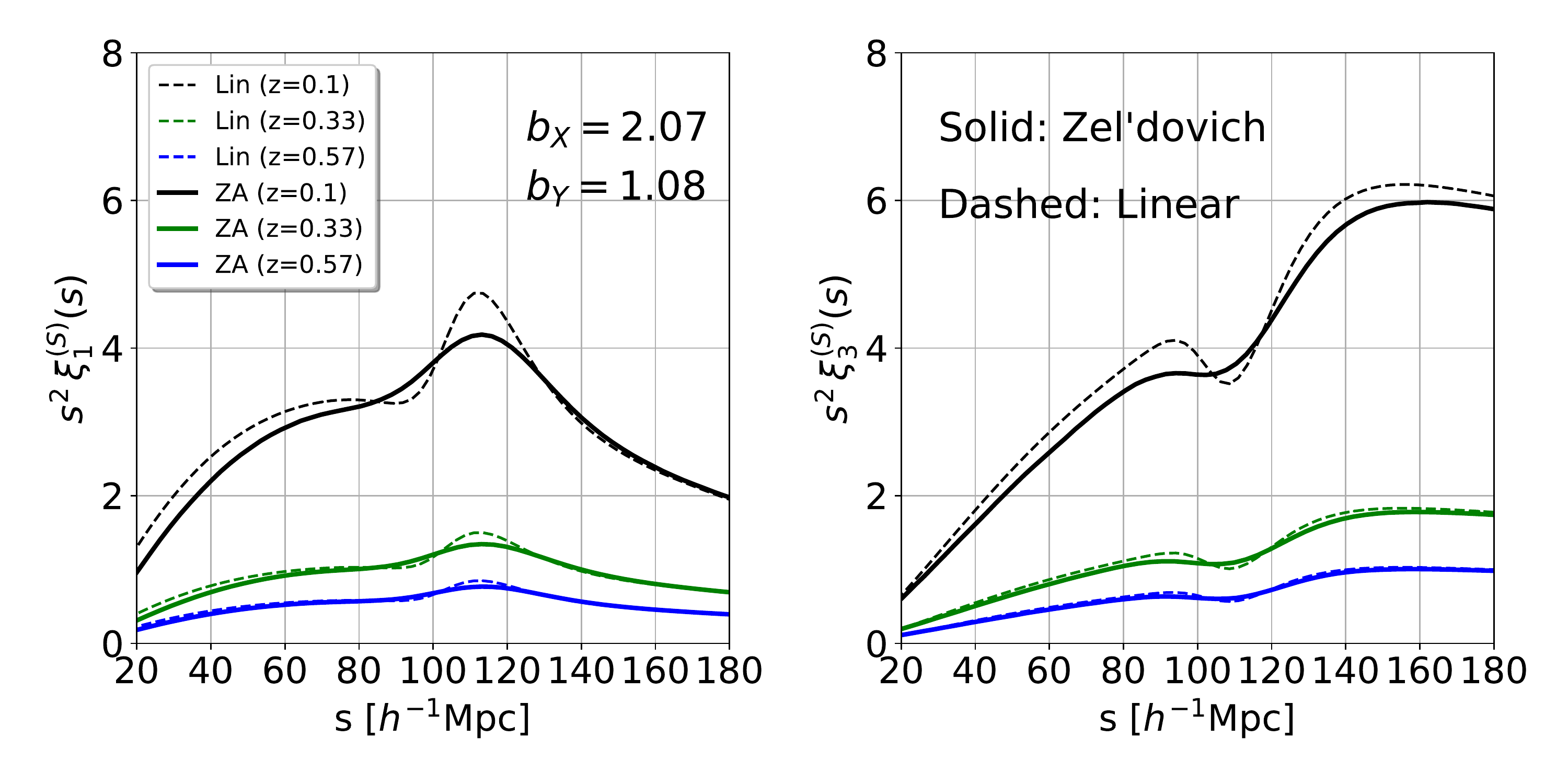}
\caption{Dipole (left) and Octupole (right) moments of cross-correlation function at $z=0.1$ (black), $0.33$ (green), and $0.57$ (blue). The plotted results are the multipole correlation function multiplied by $s^2$.  Solid and dashed lines are the predictions based on Zel'dovich approximation and linear theory, respectively. Same as in Fig.~\ref{fig:xired024_zred}, we assume the Eulerian linear bias of $b_{\rm X}=2.07$ and $b_{\rm Y}=1.08$. 
\label{fig:xired13_zred}}
\end{figure*}

One important remark of the multipole expansion in Eqs.~(\ref{eq:xired_multipole}) and (\ref{eq:xired_multipole_wideangle}) is that the wide-angle contributions crucially depends on how we choose the line-of-sight (LOS) direction, and the impact of wide-angle effects is largely changed. This point has been recently investigated in both analytical and numerical calculations \citep{Castorina_White2018b,Beutler_Castorina_Zhang2019,Reimberg_etal2016}. In what follows, keeping these aspects in mind, we will present a quantitative estimate of the impact of wide-angle effects, focusing particularly on quasi-linear scales. Analytical and numerical results presented below are obtained assuming a flat Lambda cold
dark matter (CDM) model, with the initial power spectrum created by camb \citep{Lewis:1999bs}.  
The fiducial model parameters are chosen based on the seven-year WMAP results \citep{Komatsu_etal2011}: $\Omega_{\rm m}=0.25733$ for matter density, $\Omega_{\rm b}=0.04356$ for baryon density, $\Omega_{\Lambda}=0.74259$ for dark energy with equation-of-state parameter $w=-1$, $\Omega_{\rm r}=8.076\times10^{-5}$ for radiation density, $h=0.72$ for Hubble parameter, $n_s=0.963$ for scalar spectral index, and finally, $\sigma_8=0.801$ for the normalization amplitude of the matter fluctuations at $8\,h^{-1}$Mpc.

\subsection{Deviation from plane-parallel limit}
\label{sec:departure_distant_obs}

Let us evaluate quantitatively the impact of wide-angle corrections, varying the distance to the objects, $d$. For a sufficiently long distance larger than the separation, i.e., $d\gg s$, the variation of $d$ is equivalent to that of redshift, $z$. Here, we consider the mid-point LOS as one of the simplest definitions:
\begin{align}
\mbox{Mid-point}\,:\quad \bfd\equiv \frac{1}{2}\bigl(\bfs_1 + \bfs_2\bigr).
\label{eq:mid-point_LOS}
\end{align}
Then, we compute the multipole moments of the cross-correlation function $\xi_{\ell}^{\rm(S)}$, assuming $b_{\rm X}=2.07$ and $b_{\rm Y}=1.08$ as a fiducial set of Eulerian bias parameters.

\begin{figure*}
\includegraphics[width=18cm]{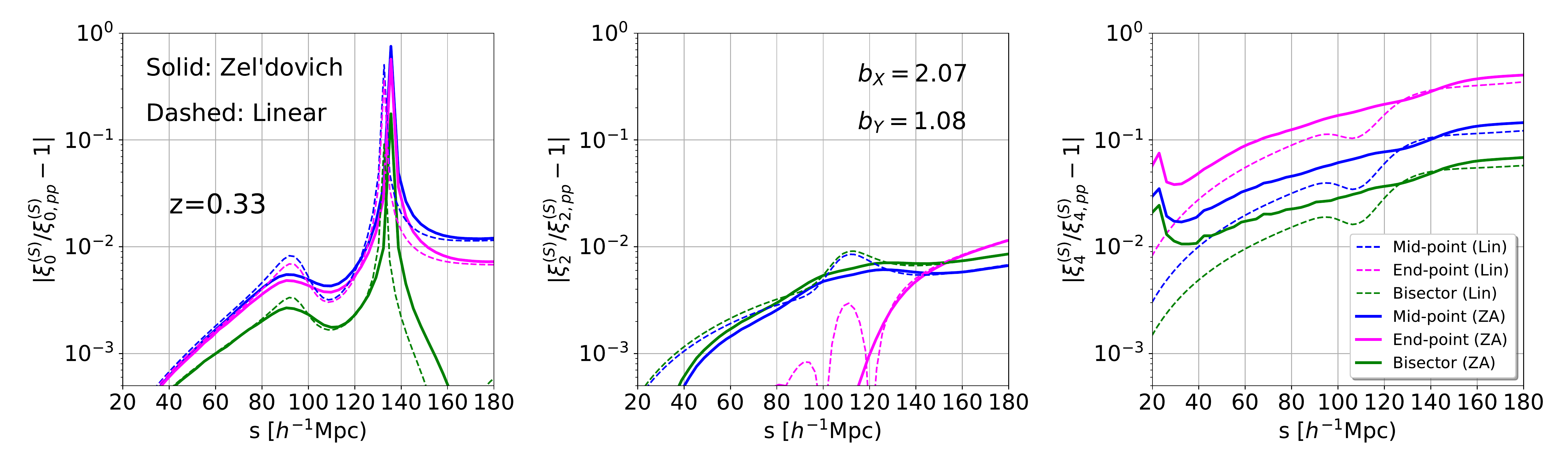}
\caption{Dependence of monopole (left), quadrupole (middle), and hexadecapole (right) cross-correlation functions on the LOS definition at $z=0.33$. Same as in Fig.~\ref{fig:xired024_zred}, we assume the Eulerian linear bias of $b_{\rm X}=2.07$ and $b_{\rm Y}=1.08$, and the fractional differences between the cross-correlation function with and without wide-angle corrections, $|\xi^{\rm(S)}_\ell(s)/\xi^{\rm(S)}_{\ell,{\rm\scriptscriptstyle pp}}(s)-1|$, are plotted in each panel. The results for the mid-point, end-point and bisector LOS are respectively shown in magenta, blue and green colors. Solid and dashed lines are the predictions based on Zel'dovich approximation and linear theory, respectively. 
\label{fig:xired024_LOS}}
\includegraphics[width=12cm]{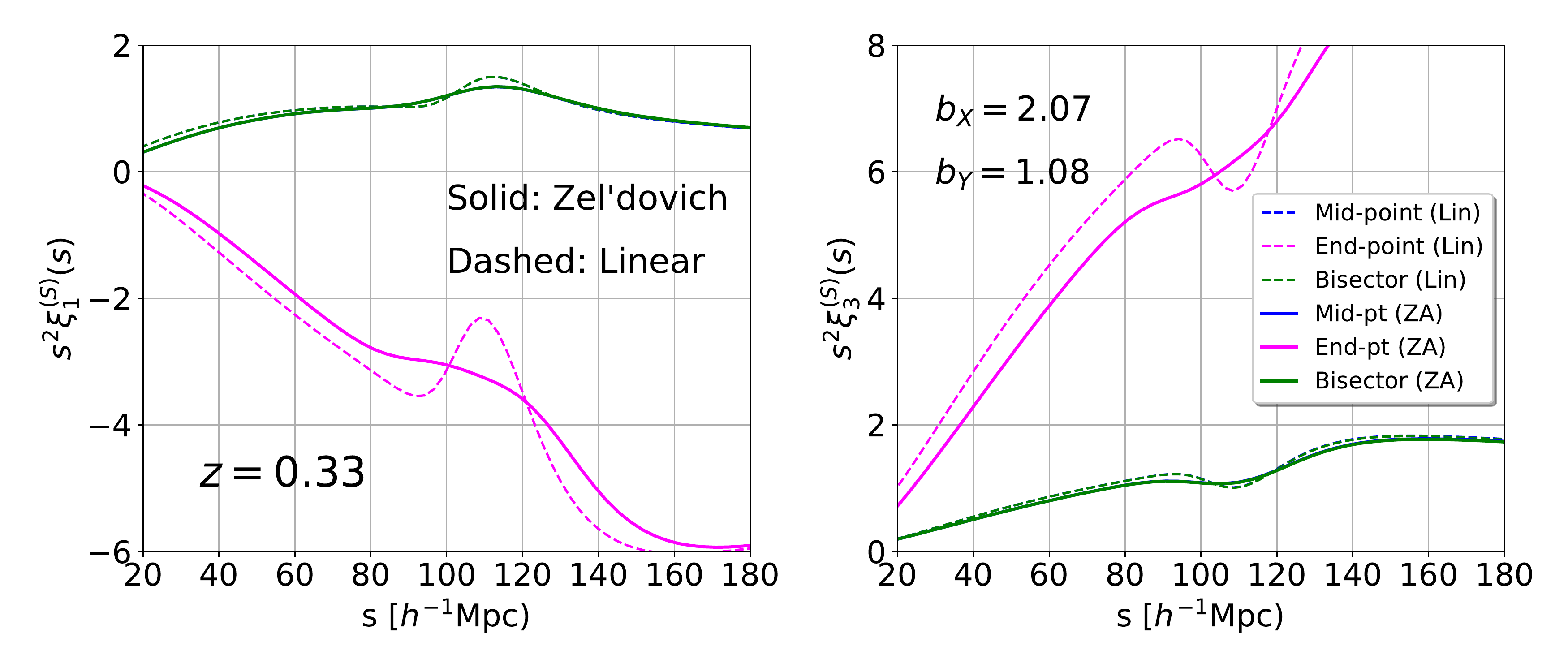}
\caption{Dependence of LOS definition on dipole (left) and octupole (right) moments of cross-correlation functions at $z=0.33$. The plotted results are the multipole correlation function multiplied by $s^2$, assuming the Eulerian linear bias of $b_{\rm X}=2.07$ and $b_{\rm Y}=1.08$. Meanings of line types and colors are the same as in Fig.~\ref{fig:xired024_LOS}. Note that the results for mid-point LOS are overlapped with those for bisector LOS. 
\label{fig:xired13_LOS}}
\end{figure*}

First look at the even multipole moments. In Fig.~\ref{fig:xired024_zred}, the results for $\ell=0$ (left), $2$ (middle), and $4$ (right), are shown at $z=0.1$ (black), $0.33$ (green), and $0.57$ (blue), corresponding to the distance $d=0.29$, $0.92$, and $1.50\,h^{-1}$\,Gpc, respectively. Note that the latter two cases are close to the mean redshifts of SDSS BOSS LRG/LOWZ and CMASS samples. The plotted quantity here is the fractional difference of the correlation function relative to the one in the plane-parallel limit, $|\xi_\ell^{\rm(S)}/\xi^{\rm(S)}_{\ell,{\rm pp}}-1|$ with $\xi^{\rm(S)}_{\ell,{\rm pp}}$ being the multipole cross-correlation function in the plane-parallel limit, equivalently $\xi_{\ell,0}^{\rm(S)}$ in Eq.~(\ref{eq:xired_multipole_wideangle}). Solid lines are the results obtained from Zel'dovich approximation, which are compared with linear theory predictions, depicted as dashed lines.

Overall, both the linear and Zel'dovich predictions give the same trend, that is, the impact of wide-angle corrections, characterized by the departure from plane-parallel limit, becomes prominent at large separation, and it is more significant at lower redshifts (small $d$). Note that a sharp feature near $s=140\,h^{-1}$\,Mpc in the monopole and $s=20\,h^{-1}$\,Mpc in the hexadecapole just comes from the zero-crossing of the correlation function. Quasi-linear prediction with Zel'dovich slightly changes the impact of wide-angle corrections in the monopole and quadrupole, and the structure of the baryon acoustic peak is smeared to some extent. The is a well-known nonlinear feature in both real and redshift space \citep[e.g.,][]{Crocce:2007dt,Matsubara2008a,Taruya:2009ir}. On the other hand, the hexadecapole exhibits a notable enhancement of the deviation from plane-parallel limit, and compared to the linear theory, it amounts to several tens of percent even at small separation. Remarkably, these behaviors are qualitatively similar to those obtained in \cite{Castorina_White2018b}, although they considered the auto-correlation function, ignoring the contributions arising from the selection function (see Figs.~3 and 4 of their paper\footnote{To be strict, Figs.~3 and 4 of their paper adopts the bisector LOS, not the mid-point LOS. Nevertheless, as we will show in Sec.~\ref{sec:LOS_definition}, the differences between bisector and mid-point LOS are sufficiently small.}). We have also examined the cases with  different values of bias parameters. Increasing $b_{\rm X}$ while keeping $b_{\rm Y}$, the resultant fractional difference is found to decrease for monopole, but to increase for hexadecapole. For quadrupole, no notable change is found. As shown in Appendix \ref{subsec:wide-angle_corrections}, the wide-angle corrections are of the order of $\mathcal{O}((s/d)^2)$, and they include the terms linearly proportional to the bias in all multipoles. Recalling the fact that in the plane-parallel limit, the monopole and quadrupole include respectively the terms proportional to $b_{\rm X}b_{\rm Y}$ and $(b_{\rm X}+b_{\rm Y})$ [see Eqs.~(\ref{eq:xi00_lin}) and (\ref{eq:xi20_lin})], the fractional difference $|\xi_\ell^{\rm(S)}/\xi^{\rm(S)}_{\ell,{\rm pp}}-1|$ tends to decrease for monopole, and to have a small bias dependence for quadrupole. On the other hand, the hexadecapole in the plane-parallel limit has no bias dependence [see Eqs.~(\ref{eq:xi40_lin})]. Thus, the fractional difference gets large as increasing the bias parameters. Although this argument is based on the linear theory formulas in \ref{subsec:wide-angle_corrections}, we expect that it generally holds even beyond linear regime.

Next look at the odd multipoles, which become vanishing in the plane-parallel limit. Any deviation from linear theory will therefore directly show up in the total signal, without taking ratio. Fig.~\ref{fig:xired13_zred} shows the dipole (left) and octupole (right) moments of cross-correlation function, multiplied by the square of separation. We see clearly the baryon acoustic feature in linear theory prediction, but it is smeared in Zel'dovich approximation, as expected from the behavior in even multipole. The amplitude of the odd multipoles is basically proportional to the difference of the bias parameter, $b_{\rm X}-b_{\rm Y}$, and hence it becomes zero in the auto-correlation case. Typically, it is smaller than that of the even multipoles by one order of magnitude. Nevertheless, it can still be detectable even with current surveys, depending on the line-of-sight definition \citep{Gaztanaga_Bonvin_Hui2017}. Since the observed relativistic effects such as gravitational redshift effect also produce non-zero odd multipoles, a quantitative prediction of odd multipoles arising from the standard Doppler effect is crucial. In this respect, the present formalism based on Zel'dovich approximation would help to disentangle several effects from the measured odd multipoles, and could be used to probe relativistic effects at quasi-linear scales.

\subsection{Dependence of line-of-sight definitions}
\label{sec:LOS_definition}

As we mentioned, the impact of wide-angle effects can change with the definition of the LOS direction. Here, we compute the cross-correlation function with several definitions of the LOS direction, and see how the results are quantitatively changed. In addition to the mid-point LOS, one may consider the end-point LOS, for which we take one of the position vectors $\bfs_1$ and $\bfs_2$ to be the LOS vector. Here, we adopt
\begin{align}
& \mbox{End-point}\,:\quad \bfd=\bfs_1.\,\,
\label{eq:end-point_LOS}
\end{align}
This definition is frequently used in measuring the multipole power spectra. One advantage of adopting Eq.~(\ref{eq:end-point_LOS}) is that one can construct a fast power spectrum estimator, making full use of the fast Fourier transform \citep{Scoccimarro2015,Bianchi_etal2015}. Another natural definition is the angular bisector line between position vectors $\bfs_1$ and $\bfs_2$: 
\begin{align}
& \mbox{Bisector}\,:\quad \bfd= \frac{s_1s_2}{s_1+s_2}(\hats_1+\hats_2). 
\label{eq:bisector_LOS}
\end{align}

In Fig.~\ref{fig:xired024_LOS}, fixing the redshift to $z=0.33$ (corresponding to the comoving distance $d=0.92\,h^{-1}$\,Gpc), we plot the fractional difference, as similarly shown in Fig.~\ref{fig:xired024_zred}, for the even multipoles. Further, in Fig.~\ref{fig:xired13_LOS}, the odd multipoles are shown, multiplying by the square of separation. Again, the cross correlation is computed assuming the linear Eulerian biases of $b_{\rm X}=2.07$ and $b_{\rm Y}=1.08$. In both figures, the results for mid-point, end-point, and bisector LOS are respectively depicted as blue, magenta and green lines. 

\begin{figure*}
\includegraphics[width=14cm]{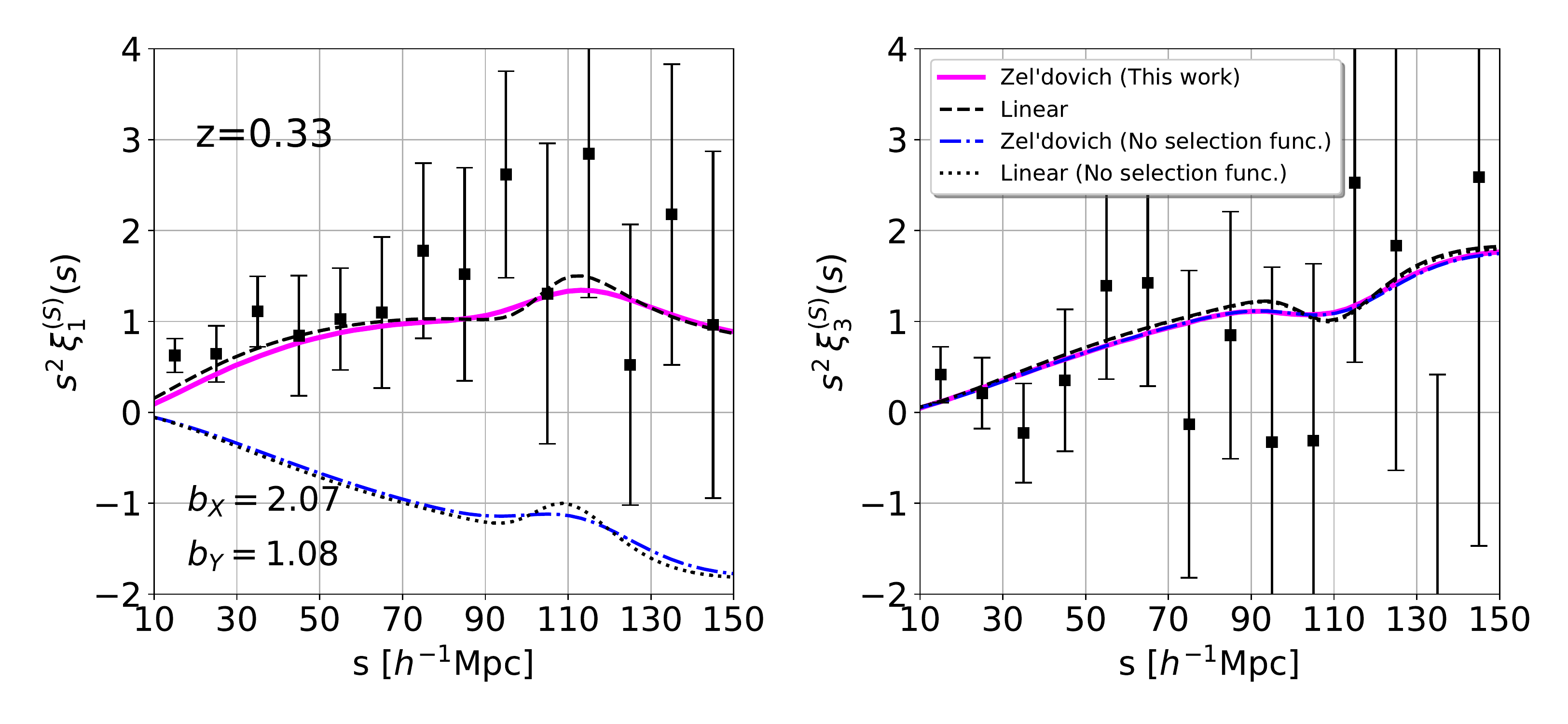}
\includegraphics[width=14cm]{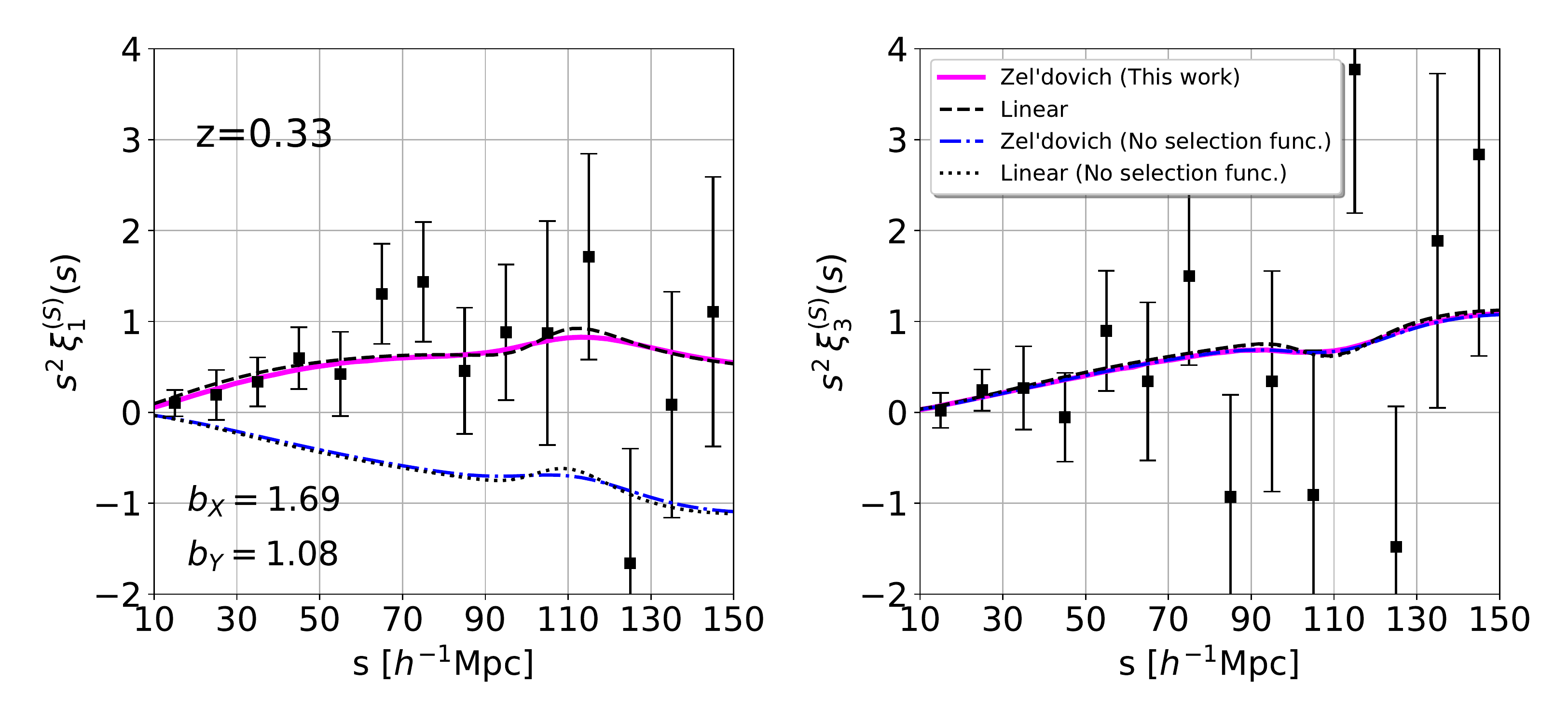}
 \caption{Comparison of the dipole (left) and octupole (right) moments of cross-correlation function between analytical predictions and measured results in $N$-body simulations, adopting the mid-point LOS given at Eq.~(\ref{eq:mid-point_LOS}). Upper and lower panels shows the results for the halos with different linear bias: $(b_{\rm X},\,b_{\rm Y})=(2.07,\,1.08)$ (upper), (1.69,\,1.08) (lower). The plotted results are the cross-correlation functions multiplied by $s^2$ at $z=0.33$. In each panel, predictions based on linear theory and Zel'dovich approximation are shown in magenta solid and black dashed lines, respectively. Also, predictions ignoring the selection function contributions are also shown in blue dot-dashed and black dotted lines, which are computed from Zel'dovich approximation and linear theory, respectively. Note that for the octupole, predictions with and without selection function contributions mostly coincide with each other.  
\label{fig:xired13_nbody}}
\end{figure*}

For even multipoles,  as we see from Fig.~\ref{fig:xired024_LOS}, the dependence of the LOS definitions is not so large except for the hexadecapole ($\ell=4$). The main reason basically comes from the fact that in linear theory, the lowest-order wide-angle corrections in Eq.~(\ref{eq:xired_multipole_wideangle}) appears at $n\geq2$ in all the three cases. In Appendix \ref{subsec:wide-angle_corrections}, we present the leading-order expressions for the wide-angle corrections to the cross-correlation in linear theory. The expressions indicate that the impact of the LOS dependence also changes with the bias parameters, but it is linear dependence on $b_{\rm X}$ and $b_{\rm Y}$. For hexadecapole, the end-point LOS definition gets a larger wide-angle correction, and for the prediction with Zel'dovich approximation, the deviation from the plane-parallel limit exceeds $10\%$ even at the scales smaller than the baryon acoustic peak, $s\gtrsim 80\,h^{-1}$\,Mpc. 

For odd multipoles, a more significant difference can be seen in Fig.~\ref{fig:xired13_LOS}. While no clear difference of the results is found for the mid-point and bisector LOS definitions, the end-point LOS gives a rather large differences in both dipole and octupole moments. In particular, the dipole correlation function changes its sign. Indeed, these behaviors are qualitatively explained by the analytic expression in linear theory, as shown in Appendix \ref{subsec:wide-angle_corrections}. At the lowest-order of expansion in Eq.~(\ref{eq:xired_multipole_wideangle}), we have
\begin{align}
 ^{\rm bisect}\xi_{1,1}^{\rm(S)}(s)&=^{\rm mid}\xi_{1,1}^{\rm(S)}(s),
\\
 ^{\rm bisect}\xi_{3,1}^{\rm(S)}(s)&=^{\rm mid}\xi_{3,1}^{\rm(S)}(s),
\end{align}
for the bisector LOS, and 
\begin{align}
 ^{\rm end}\xi_{1,1}^{\rm(S)}(s)&=^{\rm mid}\xi_{1,1}^{\rm(S)}(s)-\frac{2}{5}\,f\,\Bigl(b_{\rm X}+b_{\rm Y} + \frac{6}{7}f\Bigr)\,\Xi_2^0(s),
\\
 ^{\rm end}\xi_{3,1}^{\rm(S)}(s)&=^{\rm mid}\xi_{3,1}^{\rm(S)}(s)+
\frac{2}{5}\,f\,\Bigl(b_{\rm X}+b_{\rm Y} + \frac{6}{7}f\Bigr)\,\Xi_2^0(s)
\nonumber
\\
&\quad +\frac{16}{63}\,f^2\,\Xi_4^0(s),
\end{align}
for the end-point LOS definition. Here, the function $\Xi_m^n$ is defined by [see Eq.~(\ref{eq:def_Xi_mn_Appendix})]
\begin{align}
 \Xi_m^n (s)= \int \frac{dk\,k^2}{2\pi^2}\,\frac{j_m(ks)}{(ks)^n}\,P_{\rm L}(k).
\label{eq:def_Xi_mn}
\end{align}
Since the functions $\Xi_2^0$ and $\Xi_4^0$ give the positive contributions at the scales of our interest, the above expressions imply that the dipole moment for the end-point LOS is shown to be always smaller than that for the mid-point or bisector LOS, whereas the end-point octupole always gets a positive correction on top of the prediction for mid-point LOS. Another notable feature may be that the differences of the predictions between Zel'dovich approximation and linear theory look larger for the end-point LOS. This is particularly true at large scales beyond baryon acoustic peak. This implies that the effect of nonlinear gravitational growth would become significant for the end-point LOS, and thus an accurate nonlinear modeling would be important.

\subsection{Comparison with simulations}
\label{sec:comparison}

Finally, the predictions based on Zel'dovich approximation are compared with $N$-body simulations. For this purpose, we measure the cross-correlation functions from the full-sky halo catalog presented in \citet{Breton_etal2019}. This catalog has been created based on 
the full-sky light-cone outputs of the $\Lambda$CDM {\tt RayGalGroupSims} cosmological N-body simulation with $4,096^3$ particles in a volume of $(2.625\,h^{-1}$\,Gpc$)^3$. Using ray-tracing techniques, all the relevant relativistic contributions to the observed large-scale structure have been self-consistently incorporated under the weak-field approximation, including also the wide-angle effects on RSD. In this respect, the catalog provides an ideal suite for modeling, characterizing and testing relativistic signature detectable with future surveys. But, here, we consider the standard Doppler effect only, and rather focus on the wide-angle effects, ignoring all other contributions. A detailed comparison of the analytical model with the data including relativistic effects will be presented in a companion paper (Saga et al. in prep.). Note that the volume-averaged redshift of this catalog is $z=0.341$, and the cosmological parameters are the same as we adopted in this paper.

\subsubsection{Nonlinear impacts on odd multipoles}
\label{subsubsec:impact_nonlinear}

Fig.~\ref{fig:xired13_nbody} shows the measured results of the dipole (left) and octupole (right) cross correlations for the mid-point LOS observer, obtained from the halo sub-samples of \verb|data_H|$_{100}$ and \verb|data_H|$_{1600}$ (upper), and \verb|data_H|$_{100}$ and \verb|data_H|$_{800}$ (lower)\footnote{The label, \verb|data_H|$_N$, indicates the halo sub-sample in which each halo contains dark matter particles of the numbers ranging from $N$ to $2N$, with the mass of dark matter particle being $1.88\times 10^{10}\,h^{-1}\, M_\odot$.}. The bias of these samples is estimated from the ratio of auto-correlation function to give $1.08$, $1.69$, and $2.07$ for \verb|data_H|$_{100}$, \verb|data_H|$_{800}$ and \verb|data_H|$_{1600}$, respectively \citep[see Table 2. of][]{Breton_etal2019}. The errorbars of the measured results indicate the statistical error estimated from the jackknife method with $32$ re-samplings. Though the size of errors is large, we see clearly the non-zero signals both from dipole and octupoles, which are purely originated from the wide-angle effects. We have also measured the monopole and quadrupole moments of cross-correlation functions, which yield a much larger and clearer signal. However, they are basically dominated by the contributions from the plane-parallel limit, as shown in Figs.~\ref{fig:xired024_zred} and \ref{fig:xired024_LOS}, and hence it is difficult to isolate tiny wide-angle corrections from others.

In Fig.~\ref{fig:xired13_nbody}, analytical predictions with Zel'dovich approximation are plotted in magenta solid lines, adopting the measured linear bias parameters in \citet{Breton_etal2019} (see Table 2 of their paper). Note that in computing the cross-correlation function, we take halos with larger (smaller) bias to be the object $X$ ($Y$), so that the separation vector, given by $\bfs=\bfs_2-\bfs_1$, always points to the halos with smaller bias (see Fig.~\ref{fig:geometry_xicross}). The redshift in the analytic calculations was actually chosen to be the mean redshift of the most massive halos (i.e., \verb|data_H|$_{1600}$), $z=0.334$. To be precise, this is slightly different from the volume-averaged one ($z=0.341$), but a qualitative aspect of the comparison remains totally unchanged. In fact, the predictions agree well with measured results, and capture the overall trends, although the linear theory predictions, depicted as dashed lines, also give a good job. Since the measured odd multipoles are still noisy, one cannot clearly see that the Zel'dovich approximation outperforms the linear theory prediction. Rather, one might say that the linear theory still works well to model and predict their impacts \cite[see][for practical application]{Beutler_Castorina_Zhang2019}.

Nevertheless, as shown in \citet{Breton_etal2019}, the deviation from linear theory appears manifest when we consider the relativistic contributions. In particular, the relativistic contributions tend to have a large impact on nonlinear correction \citep[see][for a quantitative study with perturbation theory]{DiDio_Seljak2019}, and a large deviation is indeed found for the dipole purely arising from relativistic effects below $40-50\,h^{-1}$\,Mpc. In this respect, quasi-linear treatment of wide-angle effects still deserves further investigation. Extending the present formalism to include relativistic effect, we will study in detail modeling and predicting the cross-correlation functions in a separate paper (Saga et al. in prep.).

\subsubsection{Impacts of selection function contributions}
\label{subsubsec:impact_selection_func}

As a final remark, we discuss the impacts of the selection function contributions, arising from the terms proportional to $(2/s)\,(\bfv\cdot\hats)$ in linear density field [second term in Eq.~(\ref{eq:deltas_wide-angle})]. In Fig.~\ref{fig:xired13_nbody}, the predictions ignoring these contributions are plotted in blue dot-dashed and black dotted lines, which respectively indicate the results from Zel'dovich approximation and linear theory. Here, the former is obtained by tracing the the same calculation as done by \citet{Castorina_White2018b}, with a slight extension to the cross-correlation function. The latter is computed from Eq.~(\ref{eq:xistd_expansion}), taking account of a part of the coefficients $a_{mn}$ and $b_{mn}$, i.e., $a_{00}$, $a_{02}$, $a_{20}$, $a_{22}$, and $b_{22}$. 

Compared to the predictions shown in magenta solid and black dashed lines, the results in the octupole moment remain almost the same, and are hardly distinguishable from those including the selection function contributions. The fractional differences are merely $\sim2.5\%$ even at $s=200\,h^{-1}$\,Mpc, and we checked that this is also the case for even multipoles of $\ell=0$, $2$, and $4$. Remarkably, however, the dipole cross correlations exhibit a rather notable difference, with qualitatively the same trend between Zel'dovich and linear theory predictions.

These results can be deduced analytically from the linear theory based on the expansion at Eq.~(\ref{eq:xired_multipole_wideangle}) as follows. The dipole cross-correlation function for the mid-point LOS observer is expressed, at leading order, as $\xi_1^{\rm (S)}(s,d) \simeq (s/d)\, ^{\rm mid}\xi_{1,1}^{\rm(S)}(s)$ with the coefficient $^{\rm mid}\xi_{1,1}^{\rm(S)}$ given by Eq.~(\ref{eq:xi_coeff_dipole_mid}). As it has been pointed out by \citet{Bonvin_Hui_Gaztanaga2014,Tansella_etal2018,Breton_etal2019}, this coefficient can be decomposed into two parts, and the resultant dipole correlation is expressed as\footnote{The dipole contributions coming from $\xi_{\rm div}$ and $\xi_{\rm wa}$ in Eq.~(\ref{eq:xi1_div_wa}) exactly correspond to $\langle\xi^{\rm div}\rangle$ and $\langle\xi^{\rm wa}\rangle$ in \citet{Breton_etal2019} [see Eqs.~(23) and (29) of their paper].}:
\begin{align}
& \xi_{1,{\rm lin}}^{\rm(S)}(s,d)\simeq \Bigl(\frac{s}{d}\Bigr)\,\{\xi_{\rm div}(s)+\xi_{\rm wa}(s)\}\,; 
\label{eq:xi1_div_wa}
\\
 & \quad\qquad \xi_{\rm div}(s)=\frac{2}{3}\,f(b_{\rm X}-b_{\rm Y})\{\Xi_0^0(s)
+\Xi_2^0(s)\},
\nonumber
\\
 & \quad\qquad \xi_{\rm wa}(s)=-\frac{2}{5}\,f(b_{\rm X}-b_{\rm Y})\,\Xi_2^0(s).
\nonumber
\end{align}
In the above, $\xi_{\rm div}$ includes the contribution of the (uniform) selection function terms. On the other hand, the term $\xi_{\rm wa}$, which gives a negative amplitude, represents the rest of the wide-angle contributions, arising from the projection of peculiar velocities onto the radial LOS directions [third term in Eq.~(\ref{eq:deltas_wide-angle})]. That is, ignoring the selection function contributions, the predicted dipole becomes negative if $b_{\rm X}>b_{\rm Y}$. This is consistent with the results shown in Fig.~\ref{fig:xired13_nbody}. Note that for octupole, only the term corresponding to $\xi_{\rm wa}$ appears at the leading order $\mathcal{O}(s/d)$ [Eq.~(\ref{eq:xi_coeff_octupole_mid})], and thus the predictions with and without the selection functions do not show any difference.

The results suggest that a consistent framework to include all the wide-angle contributions is important in predicting the dipole, and a proper account of the selection function contribution is especially crucial. For future practical application, a further extension of the present formalism to incorporate non-uniform selection function would be important. Since a part of the relativistic effects is also known to produce similar contributions \citep[e.g.,][]{Bertacca_etal2012,Raccanelli_etal2018}, a more careful treatment may be necessary to identify and isolate the relativistic effects from others. Although the present formalism can only deal with uniform selection function, recalling the fact that the halo samples in our light-cone catalog is not perfectly uniform along the line-of-sight, a good agreement with simulations suggests some hints to approximately treat non-uniform selection function on top of the present formalism. We leave these investigations to future work.

\section{Conclusion}
\label{sec:conclusion}

The observations of large-scale structure, made through a specific observer, often break symmetries inherent in the large-scale structure. But, the symmetry breaking induced by observer can bring additional cosmological information, and offer an interesting test of cosmology. This is the redshift-space distortions (RSD) arising from the peculiar velocity of galaxies along the line-of-sight direction. Increasing the statistical precision in next-generation galaxy surveys, one will be able to not only tighten the cosmological constraints from standard RSD measurements, but also detect yet another distortion induced by the relativistic effects. In doing so, a quantitative understanding of the physical effects as well as the observational systematics is crucial issue, and a possible impacts on the cosmological interpretation needs to be investigated.

One such effect is the wide-angle effect, which appears manifest for the statistics of a widely separated galaxies. Unlike the standard RSD in which the plane-parallel limit of the observed galaxy distribution is assumed with a fixed line-of-sight direction, the translational invariance is broken for the statistical correlation of a widely separated galaxy pair, and this produces several non-trivial properties for two-point correlation function. So far, analytical study on the impact of wide-angle effects has been mostly restricted to the linear theory framework. In this paper, employing the first-order Lagrangian perturbation theory for gravitational clustering, i.e., Zel'dovich approximation, we presented a quasi-linear formalism of wide-angle effects to compute the cross-correlation function between different biased objects.

Our quasi-linear treatment of cross-correlation function is similar to what have been presented in \citet{Castorina_White2018b}. We have clarified the similarity and differences between the two treatments, and have checked in two ways that our treatment correctly reproduces the linear theory of wide-angle RSD under the uniform radial selection function. Our quasi-linear formalism with Zel'dovich approximation is thus regarded as a consistent nonlinear extension taking a proper account of wide-angle effects.

We then studied quantitatively the impact of wide-angle effects on the cross-correlation function at quasi-linear scales. In particular, we evaluated the size of the wide-angle corrections that appear in the conventional multipole expansion. We found that for even multipoles, higher multipoles tend to receive a larger wide-angle correction to the cross-correlation function, and the quasi-linear treatment with Zel'dovich approximation predicts a more significant impact of the wide-angle effects on the hexadecapole moment even at small scales. These findings are qualitatively similar to what have been found by \citet{Castorina_White2018b} in the case of the auto-correlation function. Further, a noticeable result of the cross-correlation function appears in the non-zero odd multipoles, which basically vanish in the plane-parallel limit. The amplitude of odd multipoles is roughly proportional to the difference between bias parameters, and the baryon acoustic feature is clearly seen, with the structure smeared in the quasi-linear predictions. Note cautiously that the shape of odd multipoles can be drastically changed, depending on which line-of-sight definition we use. We showed that the prediction based on the end-point line-of-sight is rather different from that for others. The linear theory formulas presented in Appendix \ref{sec:linear_theory} would provide a useful guideline to understand the line-of-sight dependence of wide-angle effects, although a quantitative understanding needs the quasi-linear treatment with Zel'dovich approximation.

Finally, we have compared our quasi-linear prediction of odd multipoles with measured results in $N$-body simulations. The predictions agree well with simulations, but within the statistical error, no noticeable difference of the predictions between linear theory and Zel'dovich approximation was found. In other words, our results implies that the linear theory description of wide-angle effects still works well at lower redshifts. Nevertheless, we note that ignoring the selection function contributions, the predicted dipole significantly deviates from simulations in both linear theory and Zel'dovich approximation. In this respect, a proper account of all wide-angle effects is crucial. Further, as it has been shown in \citet{Breton_etal2019}, when including the relativistic contributions, the linear theory prediction fails to describe the odd-multipole cross correlation at relatively large scales. A proper account of the nonlinear clustering effects seems essential for a quantitative prediction, and in this respect, the present formalism is useful and can be a basis to model and predict the cross-correlation functions including relativistic corrections. We will discuss it in more detail in a forthcoming paper.

\section*{Acknowledgments}
We are grateful to Emanuele Castorina and Martin White for their helpful and valuable comments on this paper. This work was initiated during the invitation program of JSPS Grant No. L16519. Numerical simulation was granted access to HPC resources of TGCC through allocations made by GENCI (Grand Equipement National de Calcul Intensif) under the allocations A0030402287 and A0050402287. Numerical computation was also carried out partly at the Yukawa Institute Computer Facility. This work was supported in part by MEXT/JSPS KAKENHI Grant Numbers JP15H05889 and JP16H03977 (AT). We also acknowledges the support from Grant-in-Aid  for JSPS Fellows, No. 17J10553 (SS) and No. 17J09103 (TF).



\bibliographystyle{mnras}

%
\input{references.bbl}

\appendix
\section{Analytic expression of cross-correlation function}
\label{appendix:derivation_DD_RX}

In this Appendix, starting from the expressions given at Eq.~(\ref{eq:xired_cross_DD_RR}) with (\ref{eq:R_term}) and (\ref{eq:DD_term}), 
we derive analytical expressions for cross correlation function $\xi_{\rm XY}^{\rm(S)}$ summarized at Eqs.~(\ref{eq:expression_R_part}) and (\ref{eq:expression_DD_part}), which involves three- and six-dimensional integrals for $R_{\rm X}$ and $\DD$, respectively.

\subsection{$\RX$, $\RY$-part}
\label{subsec:mean_dens}

To derive Eq.~(\ref{eq:expression_R_part}), we first make use of the fact that the quantities $\bfPsi^{\rm(S)}$ and $\deltalin$ are Gaussian fields. Then, the bracket in the integrand is rewritten with
\begin{align}
& \Bigl\langle e^{-i\bfk\cdot\bfPsi^{\rm(S)}(\bfq)} \{1+b_{\rm X}^{\rm L}\deltalin(\bfq)\bigr\}
\Bigr\rangle
\nonumber
\\
&\qquad=\exp\Bigl[-\frac{1}{2}k_ik_j\langle\Psi_i^{\rm(S)}(\bfq)\Psi_j^{\rm(S)}(\bfq)\rangle\Bigr].
\nonumber
\end{align}
Here, we used the fact that $\langle \Psi_i^{\rm(S)}(\bfq)\,\deltalin(\bfq)\rangle=0$. This implies that $R_{\rm X}=R_{\rm Y}$. Using the definition at Eq.~(\ref{eq:def_matrix_Aij}), we can rewrite Eq.~(\ref{eq:R_term}) with
\begin{align}
& \RX(\bfs)=\int\frac{d^3\bfk}{(2\pi)^3}\,\int d^3\bfq\,e^{i\,\bfk\cdot(\bfs-\bfq)}\,
\exp\Bigl[-\frac{1}{2}A_{ij}(\bfq)k_ik_j\Bigr].
\label{eq:R_X_6D-integral_form}
\end{align}
With the Gaussian integral formula at Eq.~(\ref{eq:Gaussian_forumula1}), the integral over wavevector is analytically performed to give Eq.~(\ref{eq:expression_R_part}):
\begin{align}
  \RX(\bfs)&=\int \frac{d^3\bfq}{(2\pi)^{3/2}|\mbox{det}\mbox{\boldmath$A$}|^{1/2}}\,e^{-(1/2)A^{-1}_{ij}(s-q)_i(s-q)_j}.
\nonumber
\end{align}
The explicit expression for the matrix $A_{ij}$ will be given in next subsection [see Eq.~(\ref{eq:Aij_final})].

A further reduction of the above expression is not straightforward because of the non-trivial dependence of the matrix $A_{ij}$. But, one can exploit the approximation with which $R_{\rm X,Y}$ leads to a simple analytical form. Taylor-expanding the exponential factor in Eq.~(\ref{eq:R_X_6D-integral_form}), we have
\begin{align}
R_{\rm X,Y}(\bfs)& =\int\frac{d^3\bfk}{(2\pi)^3}\,\int d^3\bfq\,e^{i\,\bfk\cdot\{\bfs-\bfq\}}\,
\sum_{n=0} \frac{1}{n!}\,\Bigl\{-\frac{k_ik_j}{2}A_{ij}(\hatq)\Bigr\}^{n}
\nonumber\\
&=\int\frac{d^3\bfk}{(2\pi)^3}\,\int d^3\bfq\,
\nonumber
\\
&\qquad\times \sum_{n=0} \frac{1}{n!}\,\Bigl\{\frac{1}{2}A_{ij}(\hatq)\frac{\partial^2}{\partial q_i\partial q_j}\,\Bigr\}^{n}
e^{i\,\bfk\cdot\{\bfs-\bfq\}},
\nonumber
\end{align}
which, repeating the integration by part, is reduced to (see also Sec.~\ref{appendix:recovery_correlation} for similar technique)
\begin{align}
R_{\rm X,Y}(\bfs)&= \sum_{n=0}\frac{1}{n!}\,\Bigl\{\frac{1}{2}\frac{\partial^2}{\partial s_i\partial s_j}A_{ij}(\hats)\Bigr\}^{n}.
\end{align}
Substituting the explicit expression of the matrix $A_{ij}$ at Eq.~(\ref{eq:Aij_final}) into the above, the approximate from of $R_{\rm X,Y}$ truncating at finite order in $A_{ij}$ is obtained to give Eq.~(\ref{eq:R_X_approx}),  
which is expressed as function of $s=|\bfs|$.

\subsection{$\DD$-part}
\label{subsec:cross_corr}

In order to derive the expression of $\DD$ relevant for numerical calculations, let us first define the following quantities:
\begin{align}
& X_1  \equiv b_{\rm X}^{\rm L}\,\deltalin(\bfq_1),
\qquad
 X_2  \equiv b_{\rm Y}^{\rm L}\,\deltalin(\bfq_2),
\nonumber
 \\
& Y \equiv -i\Bigl\{\bfk_1\cdot \bfPsi^{\rm(S)}(\bfq_1)+\bfk_2\cdot\bfPsi^{\rm(S)}(\bfq_2)\Bigr\}.
\end{align}
Then, Eq.~(\ref{eq:DD_term}) is rewritten with
\begin{align}
& \DD(\bfs_1,\bfs_2) = \int\frac{d^3\bfk_1 d^3\bfk_2}{(2\pi)^6}\int d^3\bfq_1 d^3\bfq_2\,
\nonumber
\\
&\qquad\quad \times 
e^{i\bfk_1\cdot(\bfs_1-\bfq_1)+i\bfk_2\cdot(\bfs_2-\bfq_2)}
\Bigl\langle e^Y (1+X_1)(1+X_2)\Bigr\rangle.
\label{eq:DD_term_appendix}
\end{align}
At first-order in Lagrangian perturbation theory (i.e., Zel'dovich approximation), the quantities $X_1$, $X_2$, and $Y$ all follows Gaussian statistics. Then, using the properties between moment and cumulant generating function, one can exploit the following expression \citep[see e.g.,][]{Scoccimarro:2004tg,Matsubara2008a,Taruya:2010mx}:
\begin{align}
& \Bigl\langle e^Y (1+X_1)(1+X_2)\Bigr\rangle
=\exp\Bigl[\frac{1}{2}\langle Y^2\rangle_c\Bigr]
\nonumber
\\
&\quad\times \Bigl\{1+\langle X_1X_2\rangle_c +\langle X_1Y\rangle_c + \langle X_2Y\rangle_c +\langle X_1Y\rangle_c \langle X_2Y\rangle_c \Bigr\}.
\nonumber
\end{align}
Here, the quantities enclosed by the bracket $\langle\cdots\rangle_c$ imply the cumulants, for which the disconnected part of the ensemble average is subtracted. In our case with Gaussian random fields of $X_i$ and $Y$, there is actually no distinction between cumulant and moment, and we simply omit subscript $_c$. Then, statistical quantities at right-hand side are explicitly given as follows:
\begin{align}
\langle Y^2\rangle &=-k_{1,i}k_{1,j}\,A_{ij}(\hatq_1) -k_{2,i}k_{2,j}\,A_{ij}(\hatq_2)
\nonumber
\\
&\quad -2k_{1,i}k_{2,j}\,B_{ij}(\bfq_1,\bfq_2),
\label{eq:Y2}
\\
\langle X_1Y\rangle &= i\,\bLX\,k_{2,i}\, U_i(\bfq_1,\bfq_2),
\label{eq:X1_Y}
\\
\langle X_2Y\rangle &= i\,\bLY\,k_{1,i}\,U_i(\bfq_2,\bfq_1),
\label{eq:X2_Y}
\\
\langle X_1 X_2 \rangle &= b_{\rm X}^{\rm L} b_{\rm Y}^{\rm L}\,\xi_{\rm L }(|\bfq_2-\bfq_1|),
\end{align}
where the quantity $\xi_{\rm L}$ is the correlation function of Lagrangian matter density field, $\xi_{\rm L}(|\bfq_2-\bfq_1|)\equiv\langle\deltalin(\bfq_1)\deltalin(\bfq_2)\rangle$. Here, the quantities $A_{ij}$ and $B_{ij}$ are the $3\times3$ matrices, and $U_{i}$ are the three-dimensional vectors, defined by
\begin{align}
& A_{ij}(\hatq)=\Bigl\langle \Psi_{i}^{\rm(S)}(\bfq)\Psi_{j}^{\rm(S)}(\bfq)\Bigr\rangle
\label{eq:_matrix_A1}
\\
& B_{ij}(\bfq_1,\bfq_2)=\Bigl\langle \Psi_{i}^{\rm(S)}(\bfq_1)\Psi_{j}^{\rm(S)}(\bfq_2)\Bigr\rangle
\label{eq:_matrix_B}
\\
& U_{i}(\bfq_1,\bfq_2)=-\langle \deltalin(\bfq_1)\Psi_i^{\rm(S)}(\bfq_2)\rangle.
\label{eq:_vector_U}
\end{align}
Note that Eq.~(\ref{eq:_matrix_A1}) is the same one as given at Eq.~(\ref{eq:def_matrix_Aij}). 

Substituting Eqs.~(\ref{eq:Y2})-(\ref{eq:X2_Y}) into Eq.~(\ref{eq:DD_term_appendix}), the cross correlation term becomes
\begin{align}
&\DD(\bfs_1,\bfs_2) = \int \frac{d^3\bfk_1 d^3\bfk_2}{(2\pi)^6)}\,
\int d^3\bfq_1 d^3\bfq_2\,
\nonumber
\\
&\quad\times e^{i\,\bfk_1\cdot(\bfs_1-\bfq_1)+i\,\bfk_2\cdot(\bfs_2-\bfq_2)}
\nonumber
\\
&\quad \times  \exp\Bigl[-\frac{1}{2}k_{1,i}k_{1,j}\,A_{ij}(\hatq_1)
-\frac{1}{2}k_{2,i}k_{2,j}\,A_{ij}(\hatq_2)
\nonumber
\\
&\qquad \qquad -k_{1,i}k_{2,j}\,B_{ij}(\bfq_1,\bfq_2) \Bigr]
\nonumber
\\
&\quad\times\Bigl[1+b_{\rm X}^Lb_{\rm Y}^{\rm L}\,\xi_{\rm L}(q)
+i\,\bLX k_{2,i} U_i(\bfq_1,\bfq_2)+i\,\bLY k_{1,i} U_{i}(\bfq_2,\bfq_1)
\nonumber
\\
& \qquad \qquad -\bLX\bLY\,k_{1,i}k_{2,j}\,U_i(\bfq_2,\bfq_1)\,U_j(\bfq_1,\bfq_2) \Bigr].
\label{eq:DXDY_3x3matrices_3Dvectors}
\end{align}

The above expression is further simplified if we introduce the six-dimensional vectors composed of two three-dimensional vectors, i.e., $\bfK\equiv(\bfk_1,\bfk_2)$, $\bfQ\equiv(\bfq_1,\bfq_2)$, and $\bfS\equiv(\bfs_1,\bfs_2)$. Then, Eq.~(\ref{eq:DXDY_3x3matrices_3Dvectors}) is rewritten with
\begin{align}
&\DD (\bfs_1,\bfs_2) = \int \frac{d^6\bfK}{(2\pi)^6}\,\int d^6\bfQ\,
e^{i\,K_c(S-Q)_c}
\nonumber
\\
&\quad \times \exp\Bigl[-\frac{1}{2}\mathcal{A}_{ab}(\bfQ)K_a K_b\Bigr]
\nonumber
\\
&\quad \times \Bigl[1+b_{\rm X}^{\rm L} b_{\rm Y}^{\rm L}\,\xi_{\rm L}(q)+i\,K_c\mathcal{U}_c(\bfQ)-K_aK_b\mathcal{W}_{ab}(\bfQ)\Bigr],
\end{align}
where the subscripts $a,b,c$ run over $1-6$. The quantities $\mathcal{A}_{ab}$ and $\mathcal{W}_{ab}$ are the $6\times6$ matrices, and $\mathcal{U}_a$ is the six-dimensional vector, given by
\begin{align}
\mathcal{A}_{ab} &=\left(
\begin{array}{cc}
 \bfA(\hatq_1) & \bfB(\bfq_1,\bfq_2) \\
 ^T\bfB(\bfq_1,\bfq_2) & \bfA(\hatq_2) \\
\end{array}
\right),
\label{eq:6x6_matrix_A}
\\
 \mathcal{U}_{a} &= \left(
\begin{array}{c}
 \bLY\,\bfU(\bfq_2,\bfq_1)  \\
 \bLX\,\bfU(\bfq_1,\bfq_2)  \\
\end{array}
\right),
\label{eq:6D_vector_U}
\\
\mathcal{W}_{ab}&= \frac{1}{2}\,\bLX\bLY
\nonumber
\\
&\times\left(
\begin{array}{cc}
 \bfzero & U_i(\bfq_2,\bfq_1)\,U_j(\bfq_1,\bfq_2) \\
 U_j(\bfq_2,\bfq_1)\,U_i(\bfq_1,\bfq_2) & \bfzero \\
\end{array}
\right).
\label{eq:6x6_matrix_W}
\end{align}
Now, making use of the formulas for multi-dimensional Gaussian integrals in Appendix \ref{appendix:formula_Gaussian_integrals}, the integral over the six-dimensional wavevector $\bfK$ is analytically performed, and we obtain
\begin{align}
&\DD(\bfs_1,\bfs_2) = \int \frac{d^6\bfQ}{(2\pi)^3|\mbox{det}\,\curlA|^{1/2}}\,e^{-(1/2)\mathcal{A}_{ab}^{-1}(S-Q)_a(S-Q)_b}
\nonumber
\\
&\quad \times \Bigl[1+b_{\rm X}^{\rm L}b_{\rm Y}^{\rm L}\,\xi_{\rm L}(q)
-\mathcal{A}^{-1}_{cd}\,\mathcal{U}_c(S-Q)_d 
\nonumber
\\
&\quad -\Bigl\{ \mathcal{A}^{-1}_{cd} - \mathcal{A}^{-1}_{ce}\mathcal{A}^{-1}_{df} 
(S-Q)_e (S-Q)_f \Bigr\} \mathcal{W}_{cd}
\Bigr].
\label{eq:expression_DD_part_appendix}
\end{align}
This is Eq.~(\ref{eq:expression_DD_part}). 

For a quantitative calculation of Eq.~(\ref{eq:expression_DD_part_appendix}) or (\ref{eq:expression_DD_part}), we further need explicit functional forms of $3\times3$ matrices $A_{ij}$ and $B_{ij}$ as well as thee-dimensional vectors $U_{1,i}$ and $U_{2,i}$, which are the building blocks of ${\mathcal A}_{ab}$, ${\mathcal W}_{ab}$, and ${\mathcal U}_{a}$. Recall that the displacement field in Zel'dovich approximation is related to the linear density field $\deltalin$ through Eq.~(\ref{eq:displacement_ZA}), we have
\begin{align}
 \Psi^{\rm (S)}_i(\bfq_J) & = R_{ik}(\bfq_J)\Psi_{{\rm ZA},i}(\bfq_J)
\nonumber
\\
&= R_{ik}(\bfq_J)\,\int\frac{d^3\bfp}{(2\pi)^3}\,\frac{i\,p_k}{|\bfp|^2}\,\widetilde{\deltalin}(\bfp)\,e^{i\,\bfp\cdot\bfq_J},\,\,(J=1,2)
\end{align}
with $\widetilde{\delta}_{\rm L}$ being the Fourier counterpart of the initial density field. 
Substituting the above expression into the definitions given at Eqs.~(\ref{eq:_matrix_A1})-(\ref{eq:_vector_U}), we obtain
\begin{align}
A_{ij}(\hatq_J)&=R_{ik}(\hatq_J)\, R_{jl}(\hatq_J)
\int \frac{d^3\bfp}{(2\pi)^3}\,\frac{p_kp_l}{p^2} \,P_{\rm L}(p),\,\, (J=1,2)
\label{eq:Aij_Fourier}
\\ 
B_{ij}(\bfq_1,\bfq_2)&=R_{ik}(\hatq_1)\, R_{jl}(\hatq_2)
\int \frac{d^3\bfp}{(2\pi)^3}\,\frac{p_kp_l}{p^2} 
e^{i\,\bfp\cdot(\bfq_2-\bfq_1)}\,P_{\rm L}(p),
\label{eq:Bij_Fourier}
\\ 
U_i(\bfq_1,\bfq_2)&=-R_{ik}(\hatq_2)
\int \frac{d^3\bfp}{(2\pi)^3}\,\frac{i\,p_k}{p^2} 
e^{i\,\bfp\cdot(\bfq_2-\bfq_1)}\,P_{\rm L}(p),
\label{eq:Ui_Fourier}
\end{align}
where the quantity $P_{\rm L}$ is the linear power spectrum of the density 
field $\widetilde{\deltalin}$, defined by
\begin{align}
 \langle\widetilde{\delta}_{\rm L}(\bfp)\widetilde{\delta}_{\rm L}(\bfp')\rangle=(2\pi)^3\,\delta_{\rm D}(\bfp+\bfp')\,P_{\rm L}(p).
\end{align}
Using the rotational invariance of the integrals, the above expressions are reduced to the simplified forms as 
\begin{align}
 A_{ij}(\hatq_J) &= R_{ik}(\hatq_J)R_{jk}(\hatq_J)\,\sigma_{\rm d}^2 \,;\,\quad(J=1,\,\, 2),
\label{eq:Aij_final}
\\
 B_{ij}(\bfq_1,\bfq_2) &= R_{ik}(\hatq_1)R_{jl}(\hatq_2)\Bigl\{C(q)\,\delta_{kl}+D(q)\,\hat{q}_k\hat{q}_l\Bigr\} 
\label{eq:Bij_final}
\\
 U_{i}(\bfq_1,\bfq_2) &= R_{ik}(\hatq_2)\hat{q}_k\, L(q),
\label{eq:Ui_final}
\end{align}
with $q\equiv|\bfq_2-\bfq_1|$ and $\hat{q}_k\equiv(q_{2,k}-q_{1,k})/q$. The explicit expressions for the quantity $\sigma_{\rm d}^2$ and functions $C$, $D$, and $L$ become
\begin{align}
 C(q) &=\int \frac{dp}{2\pi^2}\,\frac{j_1(pq)}{pq}\,P_{\rm L}(p),
\label{eq:func_C}
\\
 D(q) &=-\int \frac{dp}{2\pi^2}\,j_2(pq)\,P_{\rm L}(p),
\label{eq:func_D}
\\
L(q)  &= \int \frac{dp}{2\pi^2}\,p\,j_1(pq)\,P_{\rm L}(p), 
\label{eq:func_L}
\\
\sigma_{\rm d}^2 &= \int\frac{dp}{6\pi^2}\,P_{\rm L}(p),
\label{eq:sigma_d}
\\
\xi_{\rm L}(q) &= \int \frac{dp}{2\pi^2}\,p^2\,j_0(pq)\,P_{\rm L}(p),  
\label{eq:func_xiL}
\end{align}
where $j_{\ell}(x)$ is the spherical Bessel function of the first kind. 

\section{Formulas for multi-dimensional Gaussian integrals}
\label{appendix:formula_Gaussian_integrals}

Here, we summarize the formulas for Gaussian integrals used to derive the analytical expressions in Appendix \ref{appendix:derivation_DD_RX}. 
Let $K_a$ and $X_a$ be $n$-dimensional vectors, and $\mathcal{A}_{ab}$ be $n\times n$ symmetric matrix independent of $K_a$ and $X_a$. Then, we have
\begin{align}
&\int \frac{d^n\bfK}{(2\pi)^n}\,e^{i\,K_a X_a} \exp\Bigl[-\frac{1}{2}\,K_aK_b\,\mathcal{A}_{ab}\Bigr]
\nonumber 
\\
&\qquad 
=\frac{1}{(2\pi)^{n/2}|\mbox{det}\mbox{\boldmath$\mathcal{A}$}|^{1/2}}\,\exp\Bigl[-\frac{1}{2}\mathcal{A}^{-1}_{ab}X_aX_b\Bigr],
\label{eq:Gaussian_forumula1}
\\
&\int \frac{d^n\bfK}{(2\pi)^n}\,e^{i\,K_a X_a} K_c\exp\Bigl[-\frac{1}{2}\,K_aK_b\,\mathcal{A}_{ab}\Bigr]
\nonumber 
\\
&\qquad 
=\frac{i}{(2\pi)^{n/2}|\mbox{det}\mbox{\boldmath$\mathcal{A}$}|^{1/2}}\,\mathcal{A}^{-1}_{cd}\,X_d\,\exp\Bigl[-\frac{1}{2}\mathcal{A}^{-1}_{ab}X_aX_b\Bigr],
\label{eq:Gaussian_forumula2}
\\
&\int \frac{d^n\bfK}{(2\pi)^n}\,e^{i\,K_a X_a} K_cK_d\exp\Bigl[-\frac{1}{2}\,K_aK_b\,\mathcal{A}_{ab}\Bigr]
\nonumber 
\\
&\qquad 
=\frac{1}{(2\pi)^{n/2}|\mbox{det}\mbox{\boldmath$\mathcal{A}$}|^{1/2}}\,\Bigl\{\mathcal{A}^{-1}_{cd}-\mathcal{A}^{-1}_{ce}\mathcal{A}^{-1}_{df}X_eX_f\Bigr\}
\nonumber
\\
&\qquad\quad \times
\exp\Bigl[-\frac{1}{2}\mathcal{A}^{-1}_{ab}X_aX_b\Bigr].
\label{eq:Gaussian_forumula3}
\end{align}

\section{Recovery of wide-angle linear cross-correlation function}
\label{appendix:recovery_correlation}

In this Appendix, for the sake of completeness, we show that starting with the expressions involving the wide-angle effect, i.e., Eqs.~(\ref{eq:xired_cross_DD_RR}), (\ref{eq:R_term}) and (\ref{eq:DD_term}), their leading-order expansions correctly reproduce the linear theory including wide-angle effect. To do this, we keep and expand the terms up to $\mathcal{O}( P_{\rm L})$, then $\DD$, $\RX$, and $\RY$ are rewritten as
\begin{align}
&\DD(\bfs_1,\bfs_2)
\simeq \int \frac{d^3\bfk_1 d^3\bfk_2}{(2\pi)^6}\,
\int d^3\bfq_1 d^3\bfq_2\,
\nonumber
\\
&\qquad\times 
e^{i\,\bfk_1\cdot\{\bfs_1-\bfq_1\}+i\,\bfk_2\cdot\{\bfs_2-\bfq_2\}}
\,\,\Bigl[
1+ b_{\rm X}^{\rm L}b_{\rm Y}^{\rm L}\,\xi_{\rm L}(q)
\nonumber
\\
&\qquad
-\frac{1}{2}k_{1,i}k_{1,j}\,A_{ij}(\bfq_1)
-\frac{1}{2}k_{2,i}k_{2,j}\,A_{ij}(\bfq_2)
\nonumber
\\
&\qquad-k_{1,i}k_{2,j}\,B_{ij}(\bfq_1,\bfq_2)
+i\,\bLX\,k_{2,i} U_{1,i}+i\,\bLY\,k_{1,i} U_{2,i}
+\cdots 
\Bigr].
\label{eq:DXDY_linear}
\end{align}

To simplify the expression, we notice that a factor of wavenumber $\bfk_{1,2}$ in the integrand is always multiplied by the exponential $e^{i\,\bfk_1\cdot(\bfs_1-\bfq_1)+i\,\bfk_2\cdot(\bfs_2-\bfq_2)}$. Thus, we replace it with a Lagrangian spatial derivative:
\begin{align}
k_{1,i} \,\, \longrightarrow \,\,i\frac{\partial}{\partial q_{1,i}},\quad 
k_{2,i} \,\, \longrightarrow \,\,i\frac{\partial}{\partial q_{2,i}}.
\end{align}
Then, the integration can be performed analytically in a systematic manner. An explicit demonstration is given below for the term involving $B_{ij}$: 
\begin{align}
&\int \frac{d^3\bfk_1 d^3\bfk_2}{(2\pi)^6}\,
\int d^3\bfq_1 d^3\bfq_2 e^{i\,\bfk_1\cdot\{\bfs_1-\bfq_1\}+i\,\bfk_2\cdot\{\bfs_2-\bfq_2\}}
\nonumber
\\
&\qquad\qquad\qquad \times\Bigl\{
-k_{1,i}k_{2,j}\,B_{ij}(\bfq_1,\bfq_2)\Bigr\}
\nonumber
\\
&\qquad=\int \frac{d^3\bfk_1 d^3\bfk_2}{(2\pi)^6}\,
\int d^3\bfq_1 d^3\bfq_2\,
B_{ij}(\bfq_1,\bfq_2)
\nonumber
\\
&\qquad\qquad\qquad \times
\frac{\partial^2}{\partial q_{1,i}\partial q_{2,j}} 
e^{i\,\bfk_1\cdot\{\bfs_1-\bfq_1\}+i\,\bfk_2\cdot\{\bfs_2-\bfq_2\}}
\nonumber
\\
&\qquad =\int \frac{d^3\bfk_1 d^3\bfk_2}{(2\pi)^6}\,
\int d^3\bfq_1 d^3\bfq_2\,
e^{i\,\bfk_1\cdot\{\bfs_1-\bfq_1\}+i\,\bfk_2\cdot\{\bfs_2-\bfq_2\}}
\nonumber\\
&\qquad\qquad\qquad \times
\frac{\partial^2}{\partial q_{1,i}\partial q_{2,j}} 
B_{ij}(\bfq_1,\bfq_2)
\nonumber
\\
&\qquad =\int d^3\bfq_1 d^3\bfq_2\,
\delta_{\rm D}(\bfs_1-\bfq_1)\delta_{\rm D}(\bfs_2-\bfq_2)
\nonumber
\\
&\qquad \qquad\qquad 
\times \frac{\partial^2}{\partial q_{1,i}\partial q_{2,j}} 
B_{ij}(\bfq_1,\bfq_2) 
\nonumber 
\\
&\qquad =\frac{\partial^2}{\partial s_{1,i}\partial s_{2,j}} 
B_{ij}(\bfs_1,\bfs_2)
\end{align}
Note that in the third line, integration by parts is performed, assuming 
the finite support of the function $B_{ij}$. Applying the above procedure to other terms in the integrand, Eq.~(\ref{eq:DXDY_linear}) is reduced to 
\begin{align}
&\DD(\bfs_1,\bfs_2)
=
1+  b_{\rm X}^{\rm L}b_{\rm Y}^{\rm L}\,\xi_{\rm L}(s)
\nonumber
\\
&\qquad
+
\frac{1}{2}  \Bigl\{\frac{\partial^2}{\partial s_{1,i}\partial s_{1,j}}A_{ij}(\hats_1)
 + \frac{\partial^2}{\partial s_{2,i}\partial s_{2,j}} A_{ij}(\hats_{2}) \Bigr\}
\\
&\qquad 
+\frac{\partial^2}{\partial s_{1,i}\partial s_{2,j}}
    B_{ij}(\bfs_{1},\bfs_{2})
\nonumber
\\
&
\qquad
+\bLX\,\frac{\partial}{\partial s_{2,i}} U_{1,i}(\bfs_{1},\bfs_{2})
+\bLY\,\frac{\partial}{\partial s_{1,i}} U_{2,i}(\bfs_{1},\bfs_{2}).
 \label{eq:DXDY_linear2}
\end{align}
Similarly, the function $\RX$ and $\RY$ are expanded up to leading order in 
$\deltalin$, and are computed systematically to give
\begin{align}
\RX(\bfs_{1})
&= \int\frac{d^3\bfk_{1}}{(2\pi)^3}\,\int d^3\bfq_{1}\,
e^{i\,\bfk_{1}\cdot\{\bfs_{1}-\bfq_{1}\}}
\nonumber
\\
&\qquad\times\Bigl\{1
-\frac{1}{2}A_{1,ij}(\hatq_{1})k_{1,i}k_{1,j}+\cdots
 \Bigr\}
\nonumber
\\
&=
1 + \frac{1}{2}\frac{\partial^2}{\partial s_{1,i}\partial s_{1,j}}
 A_{ij}(\hats_{1}).
\label{eq:RX_linear}
\\
\RY(\bfs_{2})
&=
1 +\frac{1}{2} \frac{\partial^2}{\partial s_{2,i}\partial s_{2,j}}
A_{ij}(\hats_{2}).
\label{eq:RY_linear}
\end{align}
Combining the expressions given at Eqs.~(\ref{eq:DXDY_linear2}), (\ref{eq:RX_linear}), and (\ref{eq:RY_linear}), the leading order expression of the correlation function, $\xi_{\rm XY,lin}^{\rm(S)}$, becomes 
\begin{align}
&\xi_{\rm XY,lin}^{\rm(S)}(\bfs_1,\bfs_2)=\frac{\DD(\bfs_1,\bfs_2)}{\RX(\bfs_1)\RX(\bfs_2)} -1 
\nonumber
\\
&\quad\qquad \simeq
\bLX\bLY\,\xi_{\rm L}(q)
+\frac{\partial^2}{\partial s_{1,i}\partial s_{2,j}}
\,B_{ij}(\bfs_{1},\bfs_{2})
\nonumber
\\
&\qquad\qquad +\bLX\,\frac{\partial}{\partial s_{2,i}}
U_{1,i}(\bfs_{1},\bfs_{2})
+\bLY\,\frac{\partial}{\partial s_{1,i}}
U_{2,i}(\bfs_{1},\bfs_{2}).
\end{align}
To further reduce the above expression, we evaluate the spatial derivative of the matrix $B_{ij}$ and vectors $U_{I,i}$. Based on the expressions given at Eqs.~(\ref{eq:Bij_Fourier})-(\ref{eq:Ui_Fourier}), a straightforward calculation leads to 
\begin{align}
&\frac{\partial^2}{\partial s_{1,i}\partial s_{2,j}} B_{ij}(\bfs_1,\bfs_2)
=
\int\frac{{\rm d}^{3}\bfk}{(2\pi)^{3}}e^{i\bfk\cdot(\bfs_2-\bfs_1)}\,P_{\rm L}(k)
\nonumber
\\
&\qquad\qquad
\times \Bigl(
1 + f\mu^{2}_{1} + i\, 2 f \frac{\mu_{1}}{ks_{1}} 
\Bigr)
\Bigl(
1 + f\mu^{2}_{2} - i\, 2 f \frac{\mu_{2}}{ks_{2}} 
\Bigr), 
\label{eq: partial B}
\\
&\frac{\partial}{\partial s_{2,i}} U_{1,i}(\bfs_1,\bfs_2)
=
\int\frac{{\rm d}^{3}\bfk}{(2\pi)^{3}}e^{i\bfk\cdot(\bfs_2-\bfs_1)}\,P_{\rm L}(k)
\nonumber
\\
&\qquad \qquad \times
\Bigl(
1 + f\mu^{2}_{2}
- i\,2 f\frac{\mu_{2}}{k s_{2}}
\Bigr), 
\label{eq: partial U1} 
\\
&\frac{\partial}{\partial s_{1,i}} U_{2,i}(\bfs_1,\bfs_2)
=
\int\frac{{\rm d}^{3}\bfk}{(2\pi)^{3}}e^{i\bfk\cdot(\bfs_2-\bfs_1)}\,P_{\rm L}(k)
\nonumber
\\
&\qquad\qquad \times
\Bigl(
1 + f \mu^{2}_{1}
+ i\, 2 f\frac{\mu_{1}}{ks_{1}}
\Bigr) 
\label{eq: partial U2}
\end{align}
with $s_1=|\bfs_1|$ and $s_2=|\bfs_2|$. Here, the directional cosine $\mu_{i}$ is defined by $\mu_{i} = \hatk\cdot\hats_{i}$. Summing up the contributions above, we finally obtain
\begin{align}
&\xi_{\rm XY,lin}^{\rm(S)}(\bfs_1,\bfs_2) =
\int\frac{d^3\bfk}{(2\pi)^{3}} e^{i\bfk\cdot(\bfs_2-\bfs_1)}\, P_{L}(k)
\nonumber
\\
&\qquad\quad\times
\Bigl(
b_{\rm X} + f\mu^{2}_{1} + i\, 2 f \frac{\mu_{1}}{ks_{1}} 
\Bigr)
\Bigl(
b_{\rm Y} + f\mu^{2}_{2} - i\, 2 f \frac{\mu_{2}}{ks_{2}} 
\Bigr)
\label{eq:xired_linear}
\end{align}
with $b_{\rm X,Y}$ being the Eulerian linear bias given by $b_{\rm X,Y}=1+b_{\rm X,Y}^{\rm L}$. Eq.~(\ref{eq:xired_linear}) fully coincides with the linear theory expression that have been derived previously \citep[e.g., ][]{Papai_Szapudi2008,Yoo_Seljak2015,Reimberg_etal2016}.

\section{Linear theory of cross-correlation function with wide-angle RSD}
\label{sec:linear_theory}

In this Appendix, starting with Eq.~(\ref{eq:xired_linear}), we present the analytical formulas to compute the cross-correlation function at linear order, including the wide-angle effects.

\subsection{Expansion form of linear cross-correlation function}
\label{subsec:full_expressions}

In \citet{Szapudi2004} and \cite{Papai_Szapudi2008}, the linear-order correlation function with wide-angle effects is expanded in terms of the tripolar spherical harmonics, and it is evaluated in three different coordinate systems in the case of auto-correlation function \citep[Similar expansion has been also introduced in][]{Szalay_Matsubara_Landy1998}. Here, following \cite{Szapudi2004} and \cite{Papai_Szapudi2008}, we extend their treatment to the linear-order correlation function between different biased objects.

The tripolar spherical harmonics characterize the angular dependence of correlation function, defined by 
\begin{align}
 S_{\ell_1,\ell_2,\ell}(\hats_1,\hats_2,\hats)&=\sum_{m_1,m_2,m}
\left(
\begin{array}{ccc}
 \ell_1& \ell_2 & \ell
\\
 m_1 & m_2 & m
\end{array}
\right)
\nonumber
\\
&\quad 
\times C_{\ell_1 m_1}^*(\hats_1) C_{\ell_2 m_2}^*(\hats_2) C_{\ell m}^*(\hats)
\label{eq:def_tripolar_spherical}
\end{align}
with the function $C_{\ell m}(\hatx)$ being the normalized spherical harmonics, given by $C_{\ell m}(\hatx)\equiv \sqrt{4\pi/(2\ell+1)}\,Y_{\ell m}(\hatx)$. Note that Wigner 3$j$ symbols appear at right-hand side. With the harmonics above, we can separate the dependence of the distance and separation from their angular dependence in the cross-correlation function at Eq.~(\ref{eq:xired_linear}). We have
\begin{align}
\xi^{\rm (S)}_{\rm lin}(\bfs_1,\bfs_2)=\sum_{\ell_1,\ell_2,\ell}b_{\ell_1,\ell_2,\ell}(s_1,s_2,s)\,S_{\ell_1,\ell_2,\ell,}(\hats_1,\hats_2,\hats).
\label{eq:xi_tripolar_spherical_expansion}
\end{align}
The coefficients $b_{\ell_1,\ell_2,\ell}$ are given as the function of $s_1=|\bfs_1|$, $s_2=|\bfs_2|$, and $s=|\bfs_2-\bfs_1|$. The non-vanishing coefficients are summarized as follows:
\begin{align}
&b_{000}=\Bigl\{b_{\rm X}b_{\rm Y}+\frac{f}{3}(b_{\rm X}+b_{\rm Y})+\frac{f^2}{9}\Bigr\}\,\xi_0^2(s),
\label{eq:b000_std}
\\
&b_{220}=\,\frac{4\,f^2}{9\sqrt{5}}\,\xi_0^2(s),
\label{eq:b220_std}
\\
&b_{202}=-\frac{2\sqrt{5}}{3}\Bigl(b_{\rm Y}\,f+\frac{f^2}{3}\Bigr)\,\xi_2^2(s),
\label{eq:b202_std}
\\
&b_{022}=-\frac{2\sqrt{5}}{3}\Bigl(b_{\rm X}\,f+\frac{f^2}{3}\Bigr)\,\xi_2^2(s),
\label{eq:b022_std}
\\
&b_{222}=\frac{4}{9}\sqrt{\frac{10}{7}}\,f^2\,\xi_2^2(s),
\label{eq:b222_std}
\\
&b_{224}=4\sqrt{\frac{2}{35}}\,f^2\,\xi_4^2(s),
\label{eq:b224_std}
\\
&b_{101}=2\sqrt{3}\Bigl(\frac{b_{\rm Y}\,f}{s_1}+\frac{f^2}{3\,s_1}\Bigr)\,\xi_1^1(s),
\label{eq:b101_std}
\\
&b_{011}=-2\sqrt{3}\Bigl(\frac{b_{\rm X}\,f}{s_2}+\frac{f^2}{3\,s_2}\Bigr)\,\xi_1^1(s),
\label{eq:b011_std}
\\
&b_{121}= - 4\sqrt{\frac{2}{15}}\frac{f^2}{s_1} \xi_1^1(s),
\label{eq:b121_std}
\\
&b_{211}=  4\sqrt{\frac{2}{15}}\frac{f^2}{s_2} \xi_1^1(s),
\label{eq:b211_std}
\\
&b_{123}= - 4\sqrt{\frac{7}{15}}\frac{f^2}{s_1} \xi_3^1(s),
\label{eq:b123_std}
\\
&b_{213}=  4\sqrt{\frac{7}{15}}\frac{f^2}{s_2} \xi_3^1(s),
\label{eq:b213_std}
\\
&b_{110}=  -\frac{4\,f^2}{\sqrt{3}\,s_1 s_2} \xi_0^0(s),
\label{eq:b110_std}
\\
&b_{112}=  -4\sqrt{\frac{10}{3}}\frac{f^2}{s_1 s_2} \xi_2^0(s), 
\label{eq:b112_std}
\end{align}
where the function $\xi_\ell^m$ is defined by 
\begin{align}
  \xi_\ell^n(s)&\equiv\int\frac{dk}{2\pi^2}\,k^n\,j_{\ell}(ks)\,P_{\rm L}(k).
\label{eq:def_xi_ell_n}
\end{align}
The coefficients given above exactly coincide with those listed in \citet{Papai_Szapudi2008} if we set $b_{\rm X}=1=b_{\rm Y}$ and flip the sign for the terms involving either of factor $1/s_1$ or $1/s_2$ [see Eqs.(6)-(8) in their paper]. This is because the separation $s$, given by $s =|\bfs_2-\bfs_1|$, differs from the one defined in \citet{Papai_Szapudi2008}.

As it has been shown in \cite{Szapudi2004} and \cite{Papai_Szapudi2008}, we can further exploit a simplified expansion, which is suited for numerically computing the correlation function. To do this, based on the expansion given in Eq.~(\ref{eq:xi_tripolar_spherical_expansion}), we choose a specific coordinate system, in which the triangle formed with the position vectors $\bfs_1$ and $\bfs_2$ is confined on the $x-y$ plane, and the pair separation vector $\bfs=\bfs_2-\bfs_1$ is parallel to the $x$-axis [i.e., $\hats=(1,0,0)$]. To be precise, we set 
\begin{align}
\hats_1=\{\cos\phi_1,\sin\phi_1,0\},\quad \hats_2=\{\cos\phi_2,\sin\phi_2,0\}.
\label{eq:def_phi1_phi2}
\end{align}
This implies
\begin{align}
& S_{\ell_1,\ell_2,\ell}(\hats_1,\hats_2,\hats)
\nonumber\\
&=S_{\ell_1,\ell_2,\ell}(\{\theta_1=\pi/2,\phi_1\},\,\{\theta_2=\pi/2,\phi_2\},\,\{\theta=\pi/2,\phi=0\}).
\nonumber
\end{align}
With this choice of coordinate system, the full expressions for linear cross-correlation function can be described by a finite number of terms that depend on the two angles $\phi_1$, $\phi_2$, and distances $s_1$, $s_2$, and separation $s$: 
\begin{align}
 \xi^{\rm(S)}_{\rm XY}(\bfs_1,\bfs_2)&=\sum_{m,n}\Bigl\{a_{mn} \cos(m\phi_1)\cos(n\phi_2)
\nonumber
\\
&\quad\qquad\quad+b_{mn}\sin(m\phi_1)\sin(n\phi_2)\Bigr\}.
\label{eq:xistd_expansion}
\end{align}
The non-vanishing coefficients $a_{mn}$ and $b_{mn}$ for the linear cross-correlation function are summarized as follows:
\begin{align}
 a_{00}&=\Bigl\{b_{\rm X}b_{\rm Y}+\frac{f}{3}(b_{\rm X}+b_{\rm Y})+\frac{2\,f^2}{15}\Bigr\}\,\xi_0^2(s)
\nonumber
\\
&-\Bigl\{\frac{f}{6}(b_{\rm X}+b_{\rm Y})+\frac{2f^2}{21}\Bigr\}\,\xi_2^2(s) + \frac{3\,f^2}{140}\,\xi_4^2(s),
\nonumber
\\
 a_{02}&=-\Bigl(\frac{f}{2}\,b_{\rm X}+\frac{3\,f^2}{14}\Bigr)\,\xi_2^2(s)+\frac{f^2}{28}\,\xi_4^2(s),
\nonumber
\\
 a_{20}&=-\Bigl(\frac{f}{2}\,b_{\rm Y}+\frac{3\,f^2}{14}\Bigr)\,\xi_2^2(s)+\frac{f^2}{28}\,\xi_4^2(s),
\nonumber
\\
 a_{22}&=f^2\Bigl\{\frac{1}{15}\,\xi_0^2(s)-\frac{1}{21}\,\xi_2^2(s)+\frac{19}{140}\,\xi_4^2(s)\Bigr\},
\nonumber
\\
 b_{22}&=f^2\Bigl\{\frac{1}{15}\,\xi_0^2(s)-\frac{1}{21}\,\xi_2^2(s)-\frac{4}{35}\,\xi_4^2(s)\Bigr\},
\nonumber
\\
 a_{10}&=-\Bigl\{\frac{2\,b_{\rm Y}\,f}{s_1}+\frac{4\,f^2}{5\,s_1}\Bigr\}\,\xi_1^1(s)+\frac{f^2}{5\,s_1}\,\xi_3^1(s),
\nonumber
\\
 a_{01}&=\Bigl\{\frac{2\,b_{\rm X}\,f}{s_2}+\frac{4\,f^2}{5\,s_2}\Bigr\}\,\xi_1^1(s)-\frac{f^2}{5\,s_2}\,\xi_3^1(s),
\nonumber
\\
 a_{11}&=\frac{4\,f^2}{3\,s_1\,s_2}\Bigl\{\xi_0^0(s)-2\,\xi_2^0(s)\Bigr\},
\nonumber
\\
 a_{21}&= \frac{2\,f^2}{5\,s_2}\,\xi_1^1(s)-\frac{3\,f^2}{5\,s_2}\,\xi_3^1(s),
\nonumber
\\
 a_{12}&=-\frac{2\,f^2}{5\,s_1}\,\xi_1^1(s)+\frac{3\,f^2}{5\,s_1}\,\xi_3^1(s),
\nonumber
\\
 b_{11}&= \frac{4\,f^2}{3\,s_1\,s_2}\Bigl\{\xi_0^0(s)+\xi_2^0(s)\Bigr\},
\nonumber
\\
 b_{21}&=\frac{2\,f^2}{5\,s_2}\Bigl\{\xi_1^1(s)+\xi_3^1(s)\Bigr\},
\nonumber
\\
 b_{12}&=-\frac{2\,f^2}{5\,s_1}\Bigl\{\xi_1^1(s)+\xi_3^1(s)\Bigr\},
\nonumber
\end{align}
Note again that the coefficients $a_{mn}$ and $b_{mn}$ coincide exactly with those listed in \citet{Papai_Szapudi2008} if we set $b_{\rm X}=1=b_{\rm Y}$, and flip the sign for the terms involving either of factor $1/s_1$ or $1/s_2$.

\subsection{Line-of-sight dependent wide-angle corrections for cross-correlation function}
\label{subsec:wide-angle_corrections}

As we discussed in Sec.~\ref{sec:results}, the cross-correlation function for a widely separated pair can be also expressed as a function of line-of-sight (LOS) distance, $d=|\bfd|$, separation for a pair of objects, $s=|\bfs_2-\bfs_1|$, and the directional cosine, $\mu=\hatd\cdot\hats$, with unit vector $\hats$ defined by $\hats\equiv(\bfs_2-\bfs_2)/s$. When applying the conventional multipole expansion, we have in general the following expression [see Eqs.~(\ref{eq:xired_multipole}) and (\ref{eq:xired_multipole_wideangle})]:
\begin{align}
 \xi_{\rm XY}^{\rm(S)}(s,\,d,\,\mu)=\sum_\ell \sum_n \Bigl(\frac{s}{d}\Bigr)^n\,\xi_{\ell,n}^{\rm(S)}(s)\,\mathcal{P}_\ell(\mu). 
\label{eq:xi_wide-angle_expansion}
\end{align}
In linear theory, the leading-order expressions for the coefficients in $n$, i.e., $\xi_{\ell,0}$, are reduced to the well-known formulas in the plane-parallel limit \citep[e.g.,][]{1992ApJ385L5H}:
\begin{align}
 \xi_{0,0}^{\rm(S)}(s)&=\Bigl\{b_{\rm X}b_{\rm Y} +\frac{f}{3}(b_{\rm X}+b_{\rm Y})+ \frac{f^2}{5}\Bigr\}\,\xi_0^2(s),
\label{eq:xi00_lin}
\\
 \xi_{2,0}^{\rm(S)}(s)&=-\Bigl\{\frac{2f}{3}(b_{\rm X}+b_{\rm Y})+ \frac{4f^2}{7}\Bigr\}\,\xi_2^2(s),
\label{eq:xi20_lin}
\\
 \xi_{4,0}^{\rm(S)}(s)&=\frac{8f^2}{35}\,\xi_4^2(s).
\label{eq:xi40_lin}
\end{align}
For higher-order terms of $n\geq1$, the expressions for $\xi_{\ell,n}^{\rm(S)}$ depends on the definition of LOS direction. Below, based on the expansion form given at Eq.~(\ref{eq:xistd_expansion}), we derive the next-to-leading order expressions for the wide-angle corrections, i.e., $\xi_{\ell,1}^{\rm(S)}$ for the odd multipoles and $\xi_{\ell,2}^{\rm(S)}$ for the even multipoles, in three different definitions of LOS direction.

\subsubsection{Mid-point LOS}
\label{subsubsec:wide-angle_mid-point_LOS}

Consider first the mid-point LOS, defined at Eq.~(\ref{eq:mid-point_LOS}). With this specific definition, the position vectors for the pair of objects, $\bfs_1$ and $\bfs_2$, are expressed in terms of the LOS vector $\bfd$ and separation vector $\bfs$ as
\begin{align}
 \bfs_1=\bfd-\frac{1}{2}\bfs, \quad \bfs_2=\bfd+\frac{1}{2}\bfs.
\label{eq:s1_s2_mid-point}
\end{align}
We also recall that the two angles $\phi_1$ and $\phi_2$, defined in the specific coordinate system in Sec.~\ref{subsec:full_expressions}, are related to the position vectors $\bfs_1$  and $\bfs_2$ through Eq.~(\ref{eq:def_phi1_phi2}). With a help of these expressions and relation, we substitute the explicit form of the LOS and separation vectors, $\bfs=(s,0,0)$ and $\bfd=d(\mu,\sqrt{1-\mu^2},0)$, into the expansion at Eq.~(\ref{eq:xistd_expansion}). Then, the correlation function $\xi_{\rm XY}^{\rm(S)}$ is expressed explicitly in terms of the three variables, $s$, $d$, and $\mu$. With this expression, the correlation function is now systematically expanded in powers of $(s/d)$.

For even multipole, the next-to-leading order non-vanishing contribution to Eq.~(\ref{eq:xi_wide-angle_expansion}) appears at $n=2$. Up to $\ell=4$, we have
\begin{align}
 ^{\rm mid}\xi_{0,2}^{\rm(S)}(s)&= \Bigl\{\frac{f}{9}(b_{\rm X}+b_{\rm Y})-\frac{14\,f^2}{15}\Bigr\}\Xi_0^0(s)
\nonumber
\\
&+\Bigl\{\frac{7\,f}{90}(b_{\rm X}+b_{\rm Y})-\frac{69\,f^2}{315}\Bigr\}\,\Xi_2^0(s)+\frac{4\,f^2}{3}\,\Xi_0^2(s),
\label{eq:xi0mid2}
\\
 ^{\rm mid}\xi_{2,2}^{\rm(S)}(s)&= -\Bigl\{\frac{4\,f}{9}\,(b_{\rm X}+b_{\rm Y})+\frac{4\,f^2}{15}\Bigr\}\Xi_0^0(s)
\nonumber
\\
&-\Bigl\{\frac{23\,f}{126}(b_{\rm X}+b_{\rm Y})+\frac{23\,f^2}{147}\Bigr\}\Xi_2^0(s)-\frac{8\,f^2}{245}\,\Xi_4^0(s),
\label{eq:xi2mid2}
\\
 ^{\rm mid}\xi_{4,2}^{\rm(S)}(s)&= -\Bigl\{\frac{8\,f}{35}\,(b_{\rm X}+b_{\rm Y})+\frac{48\,f^2}{245}\Bigr\}\Xi_2^0(s)+\frac{4\,f^2}{2695}\,\Xi_4^0(s).
\label{eq:xi4mid2}
\end{align}
with the function $\Xi_m^n(s)$ defined by 
\begin{align}
 \Xi_m^n (s)\equiv \int \frac{dk\,k^2}{2\pi^2}\,\frac{j_m(ks)}{(ks)^n}\,P_{\rm L}(k).
\label{eq:def_Xi_mn_Appendix}
\end{align}
Note that this is related to the function $\xi_m^n$ at Eq.~(\ref{eq:def_xi_ell_n}) through $\Xi_m^n=\xi_m^{2-n}/s^n$. Setting $b_{\rm X}=b_{\rm Y}$, the expressions given above coincide with those obtained by \citet{Reimberg_etal2016} except for the hexadecapole, where we found a small typo in their paper [see Eqs.~(4.18)-(4.20) of their paper].

On the other hand, the odd multipoles appears non-vanishing at $n=1$. We obtain
\begin{align}
 ^{\rm mid}\xi_{1,1}^{\rm(S)}(s)&=\frac{2}{3}\,f\,(b_{\rm X}-b_{\rm Y})
\Bigl\{ \,\Xi_0^0(s) +\frac{2}{5}\,\Xi_2^0(s)\,\Bigr\},
\label{eq:xi_coeff_dipole_mid}
\\
 ^{\rm mid}\xi_{3,1}^{\rm(S)}(s)&=\frac{2}{5}\,f\,(b_{\rm X}-b_{\rm Y})
\,\Xi_2^0(s).
\label{eq:xi_coeff_octupole_mid}
\end{align}
The odd multipoles become vanishing in general for auto-correlation function (i.e., $b_{\rm X}=b_{\rm Y}$).

\subsubsection{End-point LOS}
\label{subsubsec:wide-angle_end-point_LOS}

Let us next consider the end-point LOS defined by Eq.~(\ref{eq:end-point_LOS}). In this case, the position vectors $\bfs_1$ and $\bf_2$ are expressed in terms of $\bfd$ and $\bfs$ as
\begin{align}
 \bfs_1=\bfd, \quad \bfs_2= \bfd+\bfs.
\label{eq:s1_s2_end-point}
\end{align}
Similar to the mid-point LOS case, we use Eqs.~(\ref{eq:s1_s2_end-point}) and (\ref{eq:def_phi1_phi2}) to express the expansion at Eq.~(\ref{eq:xistd_expansion}) in terms of the variables $s$, $d$, and $\mu$. 

Then, systematic expansion in power of $(s/d)$ leads to the following next-to-leading order wide-angle corrections:
\begin{align}
 ^{\rm end}\xi_{0,2}^{\rm(S)}(s)&= \Bigl\{\frac{2\,f}{9}\,b_{\rm X}-\frac{14\,f^2}{45}\Bigr\}\,\Xi_0^0(s) 
\nonumber
\\
&\quad + \Bigl\{\frac{4\,f}{45}\,b_{\rm X}-\frac{68\,f^2}{315}\Bigr\}\,\Xi_2^0(s) + \frac{4\,f^2}{3}\,\Xi_0^2(s)
\label{eq:xi0end2}
\\
 ^{\rm end}\xi_{2,2}^{\rm(S)}(s)&= -\Bigl\{ \frac{8\,f}{9}\,b_{\rm X}+\frac{4\,f^2}{15}\Bigr\}\,\Xi_0^0(s) 
\nonumber
\\
& \quad + \Bigl\{\frac{10\,f}{63}\,b_{\rm X}+\frac{10\,f^2}{147}\Bigr\}\,\Xi_2^0(s) + \frac{12\,f^2}{245}\,\Xi_4^0(s)
\label{eq:xi2end2}
\\
 ^{\rm end}\xi_{4,2}^{\rm(S)}(s)&= -\Bigl\{ \frac{32\,f}{9}\,b_{\rm X}+\frac{96\,f^2}{245}\Bigr\}\,\Xi_2^0(s) -\frac{776\,f^2}{2695}\,\Xi_4^0(s)
\label{eq:xi4end2}
\end{align}
for the even multipoles, and 
\begin{align}
 ^{\rm end}\xi_{1,1}^{\rm(S)}(s)&= \Bigl\{\frac{2\,f}{3}\,b_{\rm X}-\frac{2\,f}{3}\,b_{\rm Y}\Bigr\}\,\Xi_0^0(s)
\nonumber
\\
&\quad -\Bigl\{ \frac{2\,f}{15}\,b_{\rm X}+\frac{2\,f}{3}\,b_{\rm Y}+\frac{12\,f^2}{35}\Bigr\}\,\Xi_2^0(s)
\label{eq:xi1end1}
\\
 ^{\rm end}\xi_{3,1}^{\rm(S)}(s)&=\Bigl\{\frac{4\,f}{5}\,b_{\rm X}+\frac{12\,f^2}{35}\Bigr\}\,\Xi_2^0(s)+\frac{16\,f^2}{63}\,\Xi_4^0(s)
\label{eq:xi3end1}
\end{align}
for the odd multipoles. Note that setting $b_{\rm X}=b_{\rm Y}$ and flipping the overall sign, Eqs.~(\ref{eq:xi1end1}) and (\ref{eq:xi3end1}) coincide with those obtained by \citet{Reimberg_etal2016}\footnote{In their paper, the position vector $\bfs_2$ is taken to be the end-point LOS. }.

Note that the above expressions are related to those in the mid-point LOS case as follows:
\begin{align}
 ^{\rm end}\xi_{0,2}^{\rm(S)}(s)&=  ^{\rm mid}\xi_{0,2}^{\rm(S)}(s) 
+ \frac{f}{9}(b_{\rm X}-b_{\rm Y})\Xi_0^0(s) 
\nonumber
\\
&+ \frac{f}{90}\Bigl\{\,b_{\rm X}-7\,f-\frac{18\,f^2}{7}\Bigr\}\Xi_2^0(s),
\label{eq:xi0end2_v2}
\\
 ^{\rm end}\xi_{2,2}^{\rm(S)}(s)&=  ^{\rm mid}\xi_{2,2}^{\rm(S)}(s) 
- \frac{4\,f}{9}(b_{\rm X}-b_{\rm Y})\,\Xi_0^0(s) 
\nonumber
\\
& +\frac{43\,f}{126}\,\Bigl\{ b_{\rm X}+\frac{23}{43}\,b_{\rm Y}+\frac{198\,f}{301}\Bigr\}\Xi_2^0(s)+\frac{4\,f^2}{49}\,\Xi_4^0(s),
\label{eq:xi2end2_v2}
\\
 ^{\rm end}\xi_{4,2}^{\rm(S)}(s)&=  ^{\rm mid}\xi_{4,2}^{\rm(S)}(s) 
- \frac{24\,f}{35}\,\Bigl\{b_{\rm X}-\frac{f}{3}\,b_{\rm Y}+\frac{2\,f^2}{7}\Bigr\}\Xi_2^0(s)
\nonumber
\\
& -\frac{156\,f^2}{539}\Xi_4^0(s)
\label{eq:xi4end2_v2}
\end{align}
for even multipoles, and 
\begin{align}
 ^{\rm end}\xi_{1,1}^{\rm(S)}(s)&=^{\rm mid}\xi_{1,1}^{\rm(S)}(s)-\frac{2}{5}\,f\,\Bigl(b_{\rm X}+b_{\rm Y} + \frac{6}{7}f\Bigr)\,\Xi_2^0(s),
\label{eq:xi1end1_v2}
\\
 ^{\rm end}\xi_{3,1}^{\rm(S)}(s)&=^{\rm mid}\xi_{3,1}^{\rm(S)}(s)+
\frac{2}{5}\,f\,\Bigl(b_{\rm X}+b_{\rm Y} + \frac{6}{7}f\Bigr)\,\Xi_2^0(s)
\nonumber
\\
&\quad+\frac{16\,f^2}{63}\,\Xi_4^0(s).
\label{eq:xi3end1_v2}
\end{align}
for odd multipoles. That is, the odd multipoles for the end-point LOS generally become non-vanishing even if we set $b_{\rm X}=b_{\rm Y}$.

\subsubsection{Bisector LOS}
\label{subsubsec:wide-angle_bisector_LOS}

Finally, we consider the bisector LOS, and derive the wide-angle corrections. 
From the definition given at Eq.~(\ref{eq:bisector_LOS}) and the geometrical relation, we can express the position vectors $\bfs_1$ and $\bfs_2$ in terms of LOS vector $\bfd$ and separation vector $\bfs$ as follows:  
\begin{align}
\bfs_1 = \bfd - (1-t)\bfs, \qquad \bfs_2 = \bfd + t \bfs
\end{align}
with the quantity $t$ given by \citep{Castorina_White2018a,Castorina_White2018b}
\begin{align}
t=\frac{d+s\,\mu-\sqrt{d^2+(s\,\mu)^2}}{2\,s\, \mu}.  
\end{align}
Repeating the same procedure as given in Sec.~\ref{subsubsec:wide-angle_mid-point_LOS} and \ref{subsubsec:wide-angle_end-point_LOS},  Eq.~(\ref{eq:xistd_expansion}) is expressed in terms of the variables $s$, $d$, and $\mu$, and we can then expand it in powers of $(s/d)$.

The non-vanishing even multipoles at next-to-leading order become
\begin{align}
 ^{\rm bisect}\xi_{0,2}^{\rm(S)}(s)&=  \Bigl\{\frac{f}{9}\,b_{\rm X}+\frac{f}{9}\,b_{\rm Y} -\frac{14\,f^2}{45}\Bigr\}\,\Xi_0^0(s)  
\nonumber
\\
&\quad + \Bigl\{ \frac{f}{90}\,b_{\rm X}+\frac{f}{90}\,b_{\rm Y}-\frac{11\,f^2}{45}\Bigr\}\,\Xi_2^0(s) + \frac{4\,f^2}{3}\,\Xi_0^2(s)
\label{eq:xi0bisct2}
\\
^{\rm bisect}\xi_{0,2}^{\rm(S)}(s)&=  -\Bigl\{\frac{4\,f}{9}\,b_{\rm X}+\frac{4\,f}{9}\,b_{\rm Y} -\frac{4\,f^2}{15}\Bigr\}\,\Xi_0^0(s)  
\nonumber
\\
&\quad -\Bigl\{ \frac{29\,f}{126}\,b_{\rm X}+\frac{29\,f}{126}\,b_{\rm Y}-\frac{29\,f^2}{147}\Bigr\}\,\Xi_2^0(s) + \frac{16\,f^2}{735}\,\Xi_0^2(s)
\label{eq:xi2bisct2}
\\
^{\rm bisect}\xi_{4,2}^{\rm(S)}(s)&=  -\Bigl\{\frac{4\,f}{35}\,b_{\rm X}+\frac{4\,f}{35}\,b_{\rm Y} +\frac{24\,f^2}{245}\Bigr\}\,\Xi_2^0(s) 
\nonumber
\\
& \quad +\frac{4\,f^2}{245}\,\Xi_4^0(s),   
\label{eq:xi42bisct2}
\end{align}
which are compared with those in the mid-point LOS as follows:
\begin{align}
 ^{\rm bisect}\xi_{0,2}^{\rm(S)}(s)&=  ^{\rm mid}\xi_{0,2}^{\rm(S)}(s) -\frac{f}{15} \Bigl\{b_{\rm X}+b_{\rm Y}+\frac{6\,f}{7} \Bigr\}\Xi_2^0(s)
\label{eq:xi0bisct2_v2}
\\
 ^{\rm bisect}\xi_{2,2}^{\rm(S)}(s)&=  ^{\rm mid}\xi_{2,2}^{\rm(S)}(s) 
-\frac{f}{21} \Bigl\{b_{\rm X}+b_{\rm Y}+\frac{6\,f}{7}\Bigr\}\Xi_2^0(s)
\nonumber
\\
& +\frac{8\,f^2}{147}\,\Xi_4^0(s)
\label{eq:xi2bisct2_v2}
\\
 ^{\rm bisect}\xi_{4,2}^{\rm(S)}(s)&=  ^{\rm mid}\xi_{4,2}^{\rm(S)}(s) 
+\frac{4f}{35} \Bigl\{b_{\rm X}+b_{\rm Y}+\frac{6\,f}{7}\Bigr\}\Xi_2^0(s)
\nonumber
\\
& +\frac{8\,f^2}{539}\,\Xi_4^0(s)
\label{eq:xi42bisct2_v2}
\end{align}
Setting $b_{\rm X}=b_{\rm Y}$, Eqs.~(\ref{eq:xi0bisct2})--(\ref{eq:xi42bisct2}) are basically the same expressions as presented in \citet{Reimberg_etal2016}, where there are minor typos in Eqs.~(4.25) and (4.26).

On the other hand, the leading-order non-vanishing odd multipoles are shown to be exactly coincide with those in the mid-point LOS. That is, we have 
\begin{align}
 ^{\rm bisect}\xi_{1,1}^{\rm(S)}(s)&=^{\rm mid}\xi_{1,1}^{\rm(S)}(s),
\\
 ^{\rm bisect}\xi_{3,1}^{\rm(S)}(s)&=^{\rm mid}\xi_{3,1}^{\rm(S)}(s).
\end{align}

\bsp	
\label{lastpage}
\end{document}